\documentclass[aps,pra,amssymb, amsmath,nobibnotes,nobibnotes,reprint,superscriptaddress,showpacs,showkeys]{revtex4-1}
\usepackage[colorlinks, linkcolor=blue, anchorcolor=blue, citecolor=blue]{hyperref}
\usepackage{bm, amsfonts, mathrsfs}
\usepackage{verbatim, psfrag, times, CJK}
\usepackage{graphics, graphicx, color, epsfig}
\usepackage{float}
\usepackage{flafter,natbib}
\usepackage{indentfirst}

\begin{document}

\newcommand{\ket}[1]{| #1 \rangle}
\newcommand{\bra}[1]{\langle #1 |}

\title{First-order coherence versus entanglement in a nano-mechanical cavity}

\author{Li-hui Sun}
\altaffiliation[Also at ]{College of Physical Science and Technology, Yangtze University, Jingzhou~434023, P. R. China}
\author{Gao-xiang Li}
\email{gaox@phy.ccnu.edu.cn}
\affiliation{Department of Physics, Huazhong Normal University, Wuhan~430079,  China}
\author{Zbigniew \surname{Ficek}}
\affiliation{The National Center for Mathematics and Physics, KACST, P.O. Box 6086, Riyadh 11442, Saudi Arabia}

\date{\today}

\begin{abstract}
The coherence and correlation properties of effective bosonic modes of a nano-mechanical cavity composed of an oscillating mirror and containing an optical lattice of regularly trapped atoms are studied. The system is modelled as a three-mode system, two orthogonal polariton modes representing the coupled optical lattice and the cavity mode, and one mechanical mode representing the oscillating mirror. We examine separately the cases of two-mode and three-mode interactions which are distinguished by a suitable tuning of the mechanical mode to the polariton mode frequencies. In the two-mode case, we find that the occurrence of entanglement in the system is highly sensitive to the presence of the first-order coherence between the modes. In particular, the creation of the first-order coherence among the polariton and mechanical modes is achieved at the expense of entanglement between them. In the three-mode case, we show that no entanglement is created between the independent polariton modes if both modes are coupled to the mechanical mode by the parametric interaction. There is no entanglement between the polaritons even if the oscillating mirror is damped by a squeezed vacuum field. The interaction creates the first-order coherence between the polaritons and the degree of coherence can, in principle, be as large as unity. This demonstrates that the oscillating mirror can establish the first-order coherence between two independent thermal modes. A further analysis shows that two independent thermal modes can be made entangled in the system only when one of the modes is coupled to the intermediate mode by a parametric interaction and the other is coupled by a linear-mixing interaction.
\end{abstract}


\maketitle

\section{Introduction}

The generation of quantum effects in nano-mechanical cavities with movable mirrors has been the subject of a great interest in recent years~\cite{gm09,agh10}. This interest stems from the possibility of the development of new practical techniques for cooling macroscopic objects to very low temperatures and for engineering of entangled states of macroscopic systems. With the recent progress in laser cooling techniques, fabrication of low-loss optical elements and high-$Q$ mechanical resonators, it is now possible to prepare nanomechanical oscillators that can be controlled to a very hight precision and can even reach the quantum level of the oscillations~\cite{td09}. In these systems, the vibrations of mechanical oscillators are induced by radiation pressure that creates a strong nonlinear coupling of the vibrational mode to radiation modes. Most of these studies have been done on examples provided by cavity optomechanical systems with linear or ring cavities~\cite{gh09}. It has been demonstrated that a large radiation pressure can be generated in the cavity that in return may lead to entanglement between different components of the system. In particular, stationary entanglement has been predicted between the cavity mode and a vibrating mirror~\cite{fg06,vg07,pv07,vt07,bg08}, between an atomic ensemble or a Bose-Einstein condensate located inside an optical cavity and the vibrating mirror of the cavity~\cite{gm08,ig08,pc10,cp11}, between two vibrating mirrors of a ring cavity~\cite{mg02}, between two dielectric membranes suspended inside a cavity~\cite{hp08}, and between a membrane and a trapped atom both located inside a cavity~\cite{hw09,wh10,ck11}. Further studies have addressed interesting problems of entangling mechanical oscillators~\cite{jh09}, entangling optical and microwave cavity modes~\cite{bv11}, and the creation of a photon by a vibrating mirror~\cite{ak10}. In this connection, we should mention the most recent work on the generation of entanglement in pulsed cavity optomechanics~\cite{hw11}, and the work on the creation of entanglement between two oscillating mirrors through the coupling of the mirrors to an atomic system~\cite{zh11}.

The purpose of this paper is to explore coherence and correlation features of an optomechanical system composed of three bosonic modes realized with an one-dimensional optical lattice located inside a single-mode cavity with a movable mirror. We are particularly interested in the effect of the first-order correlation (first-order interference) on entanglement between two modes which is associated with second-order correlation functions. It is well known that in the parametric down-conversion, served as a typical source of entanglement, the signal and the idler beams are strongly entangled but always behave as mutually incoherent~\cite{gh86,ow90,mg96}. A similar conclusion applies to the correlations between modes of the optomechanical system, where it was demonstrated that the cavity and mechanical modes play the same role as the signal and the idler of a nondegenerate parametric oscillator and the modes behave as mutually incoherent~\cite{vt07,ig08}. This seems to suggest that entanglement between two modes rules out the first-order coherence between them.

We consider various situations where the modes of the optomechanical system can be made mutually coherent and to exhibit the first-order interference. This leads to an obvious question of to what extent of the first-order coherence could affect entanglement between the modes. To address this question, we use a polariton model of the optical lattice coupled to a single-mode cavity field and calculate various coherence and correlation functions of the optomechanical system. We show that the system is capable of generating a wide class of coherence and correlation effects, ranging from the first-order coherence, the anomalous autocorrelations and anomalous cross correlations between the modes. In a series of simple examples we show that the generation of the first-order coherence between two modes of the system is equally effective in destroying entanglement between these modes. We illustrate our considerations by examining two and three mode interactions. After establishing the connection between the generation of entanglement and the first-order coherence, we consider the problem of the creation of entanglement between two independent modes by coupling them to an intermediate mode. We show that the coupling of the modes to the intermediate mode by a  parametric interaction results in no entanglement between the modes. We then consider a different coupling configuration and find that an entanglement can be generated between two independent modes if one of the modes is coupled to the intermediate mode by the parametric interaction and the other is coupled by the linear-mixing interaction.

The paper is organized as follows. We begin in Sec.~\ref{sec2} with the description of the model. We represent the finite size optical lattice that is located inside a single-mode cavity in terms of Bloch-type waves called excitons, and diagonalize the interaction Hamiltonian of the excitons plus the cavity mode to describe the system in terms of bright polaritons. We then proceed in Sec.~\ref{sec3} to study the dynamics of the system in terms of the quantum Langevin equations for relevant variables. We use the linearization approach to the equations of motion and arrive to a set of three couple differential equations for the fluctuation operators, which we solve for the steady-state. We apply the solution in Sec.~\ref{sec4} to the calculation of the bipartite coherence and correlation functions of the polariton and the mechanical modes. We discuss the conditions for entanglement in terms of the squeezing fluctuations and the Cauchy-Schwartz inequality. The anomalous autocorrelation and cross correlation functions are introduced to discuss conditions for the violation of the Cauchy-Schwartz inequality. A possibility of generating a two-color entanglement is also discussed. In Sec.~\ref{sec5} we evaluate the first-order coherence and entanglement in the case of three coupled modes. Two coupling configurations of two independent modes to the intermediate mode are discussed. In Sec.~\ref{sec6}, we examine parameter ranges in which the predicted coherence and correlation effects could be observed with the current experiments. Finally, in Sec.~\ref{sec7} we summarize our results.

\section{The model}~\label{sec2}

\begin{figure}[thb]
\begin{center}
\begin{tabular}{c}
\includegraphics[width=0.9\columnwidth]{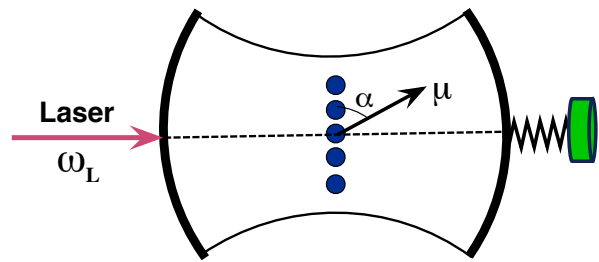}
\end{tabular}
\end{center}
\caption{(Color online) Schematic diagram of the system. An optical lattice composed of regularly spaced atoms is located inside a single-mode cavity driven by a laser field. The cavity is composed of one fixed and one movable mirror that can undergo harmonic oscillations due to the radiation pressure induced by  the laser field. The atomic sites have parallel dipole moments $\vec{\mu}$ oriented in the direction making the angle~$\alpha$ with the lattice direction.}
\label{mirfig1}
\end{figure}

We consider a finite size one-dimensional optical lattice located inside a single-mode cavity with one fixed partially transmitting mirror and one movable perfectly reflecting mirror, as shown in Fig.~\ref{mirfig1}. The optical lattice is composed of~$N$ regularly spaced  and non-overlapping sites located at positions $r_{n} = nd, \, n=1,\ldots ,N$, with total length $L = Nd \ll w_{0}$, where $d$ is the separation between the sites and $w_{0}$ is the cavity mode waist at the position of the lattice~\cite{zr09}. The lattice is formed by two counterpropagating laser beams entering the cavity from the sides and forming a standing wave in the direction perpendicular to the direction of the cavity mode. The cold atoms loaded on the optical lattice are confined in an array of microscopic trapping potentials, forming a Mott-insulator-like medium with one atom per site~\cite{zr07,jb98}.

The atoms in the optical lattice are modeled as two-level systems with ground state $\ket{g_{n}}$ and excited state $\ket{e_{n}}$, separated by the transition frequency $\omega_{a}$ and connected by a transition dipole moment $\vec{\mu}=\bra{e_{n}}\vec{\mu}_{n}\ket{g_{n}}$, which can be assumed to be real with no loss of generality. Since the optical lattice is formed in the direction perpendicular to the cavity axis, the effect of the motion of the atoms (center of mass motion) on the coupling strength of the atoms to the cavity mode and on the radiation pressure on the movable mirror can be ignored. The situation would be different and the center of mass motion important if the optical lattice were generated along the cavity axis by the cavity mode~\cite{mm06}, or by a standing wave formed by running and reflected from the movable mirror laser beams~\cite{hs10}.

The motion of the movable mirror is modeled as a quantum mechanical harmonic oscillator of mass $m$ and resonant frequency $\omega_{m}$. The cavity mode is driven by an external laser field which is treated classically in our calculations and is characterized by its frequency $\omega_{L}$ and amplitude $E_{L}$. It has become common to consider the laser field as a source of the radiation pressure force on the movable mirror.

The total Hamiltonian $H$ of the system can be written as
\begin{align}
H = H_{c}+H_{ex}+H_{0} +H_{I} ,\label{e1}
\end{align}
where
\begin{align}
H_{c}=\hbar \omega_c a^\dag a \label{e2}
\end{align}
is the free Hamiltonian of the cavity mode
\begin{align}
H_{ex} = \hbar\sum_n \omega_{a} B^\dag_n B_n +\hbar\sum_{n\neq m}J_\alpha B^\dag_n B_m ,\label{e3}
\end{align}
is the Hamiltonian of the electronic excitation in the atoms of the optical lattice
\begin{align}
H_{0} = \frac{1}{2}\hbar \omega_{m}\left(q^2+p^2\right) ,\label{e4}
\end{align}
is the free Hamiltonian of the oscillating mirror, and
\begin{align}
H_{I} &= \hbar\sum_{n}g_{n}\left(B^\dag_n a + a^{\dagger}B_n\right)  -\hbar G_{0} a^{\dag} a q \nonumber\\
&+ \hbar \left(E_{L}a^\dag {\rm e}^{-i\omega_L t} + E_{L}^{\ast}a {\rm e}^{i\omega_L t}\right)  \label{e5}
\end{align}
is the interaction Hamiltonian of the cavity mode with the electronic excitations, the movable mirror and the external laser field, respectively.

Here, $a^{\dag}$ and $a$ are creation and annihilation operators of the cavity mode of frequency $\omega_{c}$. The operators $q$ and $p$ are, respectively, the dimensionless position and momentum operators of the oscillating mirror that satisfy the fundamental commutation relation $[q,p]=i$. The contribution of electronic excitations in atoms is expressed as a sum over normal Boson creation and annihilation operators, $B^{\dag}_{n}$ and $B_{n}$, respectively, one for each site $n$ of energy $\hbar\omega_{a}$, and
\begin{eqnarray}
J_\alpha =\frac{\mu^2}{4 \pi \epsilon_0 \hbar d^{3} }\left(1-3\cos^2\alpha\right)
\end{eqnarray}
represents the contribution of the nearest-neighbor dipole-dipole interaction between atomic sites, with parallel dipole moments oriented in the direction making the angle $\alpha$ with the lattice direction. In a practical situation, the angle $\alpha$  is fixed by the polarization direction of the cavity mode. At low number of electronic excitations, that we consider here, the operators $B_n$ can be treated as bosonic operators.

The first term in the Hamiltonian (\ref{e5}) describes the interaction of the electronic excitations with the cavity field. The strength of the interaction is characterized by the Rabi frequency $g_{n}$. The interaction retains only the terms which play a dominant role in the electric-dipole and rotating-wave approximations. The higher order and antiresonant terms which would make much smaller contributions have been omitted.

The second term in the Hamiltonian (\ref{e5}) describes the optomechanical radiation-pressure interaction which couples the cavity photon number $n_{c}=a^{\dag}a$ to the position operator $q$ of the oscillating mirror with the coupling strength $G_{0}=(\omega_c/L_{c})\sqrt{\hbar/m \omega_m}$, where $m$ is the effective mass of the mechanical mode and $L_{c}$ is the length of the cavity. Finally, the parameter $E_{L}$ describes the coupling strength of the laser field to the cavity mode.

We now consider the energy states of the optical lattice that is determined by the Hamiltonian (\ref{e3}). It is easily verified that in the bare basis of the lattice sites, $\{\ket{U_{1}},\ket{U_{2}},\ldots ,\ket{U_{N}}\}$, where $\ket{U_{i}} =\ket{e_{i}}\prod_{j\neq i} \ket{g_{j}}$, the Hamiltonian (\ref{e3}) is not diagonal due to the presence of the dipole-dipole interaction $J_{\alpha}$. The diagonalization of the Hamiltonian (\ref{e3}) results in  eigenstates described by Bloch-type waves, called the collective excitation modes or shortly excitons. The diagonal Hamiltonian is of the form
\begin{eqnarray}
H_{ex} =\hbar \sum_k \omega_k C^\dag_k C_k ,
\end{eqnarray}
where
\begin{eqnarray}
\omega_{k} = \omega_{a} + 2J_{\alpha} \cos\left(\frac{\pi k}{N+1}\right)  \label{e12}
\end{eqnarray}
is the frequency of the $k$th exciton mode, and $C_{k}^{\dag}$ and $C_{k}$ are, respectively the creation and annihilation operators of the excitons. The operators $C_{k}^{\dag}$ and~$C_{k}$ are obtained from the creation and annihilation operators of an electronic excitation at site $n$ using the transformation
\begin{eqnarray}
B_n=\sqrt{\frac{2}{N+1}}\sum_k \sin\left(\frac{\pi n k}{N+1}\right)C_k .
\end{eqnarray}
Simply, the $C_{k}$ operators are obtained by inverting the above transformation.

Thus, in terms of the exciton operators, the total Hamiltonian of the system can be written in the form
\begin{align}
H = H_{0}+H_{I} ,\label{e14}
\end{align}
where $H_{0}$ is the free Hamiltonian
\begin{align}
H_{0} &=\hbar \omega_c a^\dag a+\sum\limits_{k}\hbar \omega_{k} C_{k}^\dag C_{k}
+\frac{1}{2}\hbar \omega_m \left(q^2+p^2\right) ,\label{e15}
\end{align}
and $H_{I}$ is the interaction Hamiltonian
\begin{align}
H_{I} &= \hbar \sum_{odd \ k}f_{k}\left(C_{k}a^\dag + {\rm H.c.}\right) -\hbar G_{0} a^{\dag} a q \nonumber\\
&+\hbar \left(E_{L}a^\dag {\rm e}^{-i\omega_L t}  + E_{L}^{\ast}a{\rm e}^{i\omega_L t}\right) ,\label{e19}
\end{align}
with
\begin{eqnarray}
f_{k} = \sqrt{\frac{\omega_c \mu^2}{\hbar\epsilon_0 V (N+1)}}\cot\left[\frac{\pi k}{2(N+1)}\right] .\label{e20}
\end{eqnarray}
Note that the interaction involves exciton modes with odd $k$ only~\cite{zr09}. In addition, the coupling constants $f_{k}$ are not identical, so that the exciton modes $k$ are not equally coupled to the cavity mode. Since the cotangent function decreases rapidly with $k$, one can easily verified that the strength of the coupling of the $k=1$ mode is stronger by $k^{2}$ from the $k\neq 1$ modes. This indicates that only the $k=1$ mode can be strongly coupled to the cavity field, with the modes $k\neq 1$ weakly coupled to the field.

The strong coupling of the $k=1$ exciton to the cavity mode prompts us to write the Hamiltonian in the form
\begin{align}
H = H_{1}+H_{2} ,\label{e20a}
\end{align}
where
\begin{align}
H_{1} &=\hbar \omega_c a^\dag a+\hbar \omega_{1} C_{1}^\dag C_{1}
+\hbar f_{1}\left(C_{1}a^\dag + {\rm H.c.}\right) ,\label{e20b}
\end{align}
and
\begin{align}
H_{2} &= \frac{1}{2}\hbar \omega_m \left(q^2+p^2\right)-\hbar G_{0} a^{\dag} a q \nonumber\\
&+ \hbar\left(E_{L}a^\dag {\rm e}^{-i\omega_L t} + E_{L}^{\ast}a {\rm e}^{i\omega_L t}\right) .\label{e20c}
\end{align}
We may diagonalize the Hamiltonian $H_{1}$ to find new operators of the combined $k=1$ exciton plus the cavity field system. The cavity mode can be considered to "dress" the exciton and to form along with it a single "polariton" quantum system. This reflects the fact that photons are exchanged between the exciton and cavity modes. The dressed operators, the eigen-operators of the Hamiltonian $H_{1}$, are found by the following unitary transformation
\begin{align}
\Psi &= (\cos\phi)C_{1} - (\sin\phi)a ,\nonumber\\
\Phi &= (\sin\phi)C_{1}  + (\cos\phi)a  ,\label{e20dd}
\end{align}
where the rotation angle $\phi$ is defined by
\begin{align}
\cos^{2}\phi = \frac{1}{2} +\frac{\delta}{2\Omega} ,
\end{align}
with $\delta = (\omega_{c}-\omega_{1})/2$ and $\Omega = (f_{1}^{2} +\delta^{2})^{\frac{1}{2}}$.
The angle~$\phi$ belongs to the interval $[0,\pi/2]$. The polaritons are coherent superpositions of the exciton and the cavity fields. For $\delta =0$, these are equally weighted, maximally entangled, superpositions of the fields, whereas for a large positive detuning, $\delta \gg f_{1}$, the exciton and the cavity field disentangle that then then polariton $\Psi$ becomes purely atomic (excitonic), while the polariton $\Phi$ becomes purely photonic.

In terms of the polariton creation and annihilation operators, the Hamiltonian (\ref{e20a}) takes the  form
\begin{align}
H &= \hbar (\omega_{0} -\Omega)\Psi^{\dag}\Psi + \hbar (\omega_{0} +\Omega)\Phi^{\dag}\Phi\nonumber\\
& +\frac{1}{2}\hbar \omega_m \left(q^2+p^2\right) -\hbar G_{0} a^{\dag} a q \nonumber\\
&+ \hbar \left(E_{L}a^\dag {\rm e}^{-i\omega_L t} + E_{L}^{\ast}a {\rm e}^{i\omega_L t}\right) ,\label{e27}
\end{align}
where $\omega_{0}=(\omega_{c}+\omega_{1})/2$ is the mid-frequency of the two polariton modes, and the annihilation operator for the cavity mode is related to the annihilation operators of the polariton modes by $a = \Phi\cos\phi - \Psi\sin\phi$.

This shows the familiar coupled exciton-photon mode splitting~\cite{wn92}. The mid-frequency of the polariton modes is an average value of the cavity frequency and the excitonic frequency. Thus, $\omega_{0}$ is always between $\omega_{c}$ and $\omega_{1}$, that the frequency $\omega_{0}$ is pulled away from the cavity frequency towards the excitonic frequency. When the cavity frequency is tuned to exact resonance with the excitonic frequency, i.e., $\omega_{c}=\omega_{1}$, there is no mode pulling, i.e. $\omega_{0}=\omega_{c}$.

\section{Linearized fluctuation analysis}~\label{sec3}

Given the Hamiltonian of the system, we now proceed to study the dynamics of the system in terms of the Heisenberg equations of motion for relevant variables, the polariton modes and the mirror mode operators. However, a proper analysis of the system must include losses due to the coupling of the cavity mode, the polaritons and the oscillating mirror to their local environments. Therefore, we introduce so-called phenomenological damping terms that the cavity field amplitude is damped with a rate $\kappa$, the amplitude of the exciton's field is damped with the atomic spontaneous emission rate~$\gamma_{a}$, and the oscillations of the cavity mirror are affected by the quantum Brownian noise acting on the mirror leading to the damping of its oscillations with a rate $\gamma_{m}$. The inclusion of the losses to the Heisenberg equations of motion results in a set of nonlinear Langevin equations, that written in the frame rotating at the laser frequency $\omega_{L}$ have the form
\begin{align}
\dot{q} &= \omega_m p  ,\nonumber\\
\dot{p} &= -\gamma_m p -\omega_m q+G_{s}\Psi^{\dag}\Psi +G_{c}\Phi^{\dag}\Phi \nonumber\\
&-\frac{1}{2}G\left(\Psi^{\dag}\Phi +\Phi^{\dag}\Psi\right) +\sqrt{2\gamma_{m}}\xi ,\nonumber\\
\dot{\Psi} &= -E_{L}^{\ast}\sin\phi -\left[\gamma+i\left(\Delta_{L} -\Omega\right)\right]\Psi +iG_{s}q \Psi  \nonumber\\
&-\frac{1}{2}iGq\Phi +\sqrt{2\gamma}\Psi_{in} ,\nonumber\\
\dot{\Phi} &= E_{L}^{\ast}\cos\phi -\left[\gamma+i\left(\Delta_{L} +\Omega\right)\right]\Phi +iG_{c}q \Phi  \nonumber\\
&-\frac{1}{2}iGq\Psi +\sqrt{2\gamma}\Phi_{in} ,\label{e21}
\end{align}
along with the corresponding equations for the adjoint operators $\Psi^{\dag}$ and $\Phi^{\dag}$. Here, $\Delta_{L} =\omega_{0}-\omega_{L}, G_{s}=G_{0}\sin^{2}\phi, G_{c}=G_{0}\cos^{2}\phi$, $G=G_{0}\sin(2\phi)$, and we have chosen $\gamma_{a}=\kappa =\gamma$.  Moreover, the laser field is assumed to be tuned close to the cavity and atomic resonance, in the sense that the detuning $\Delta_{L}$ is small compared to the optical frequency $\omega_{L}$.
We shall assume additionally that the input noises to the polariton modes,~$\Psi_{in}$ and $\Phi_{in}$, which are the sum of the input noises to the cavity and to the excition modes, are frequency independent Gaussian (white) vacuum noises, so that all first moments vanish, $\langle \Psi_{in}\rangle = \langle \Psi_{in}^{\dag}\rangle = \langle \Phi_{in}\rangle = \langle \Phi^{\dag}_{in}\rangle \equiv 0$, and only non-zero are the following second moments
\begin{align}
\langle \Psi_{in}(t) \Psi^\dag_{in}(t^\prime)\rangle = \langle \Phi_{in}(t) \Phi^\dag_{in}(t^\prime)\rangle = \delta(t-t^\prime).\label{e22}
\end{align}
Similarly, for the quantum Brownian noise $\xi(t)$ which arises from the coupling of the oscillating mirror to its local environment, we assume that it is a frequency independent Gaussian thermal white noise, so that all first moments vanish, $\langle \xi(t)\rangle \equiv 0$, and the only non-zero are the following second moments
\begin{align}
\frac{1}{2}\left(\langle\xi(t)\xi(t^\prime)\rangle +\langle\xi(t^\prime)\xi(t)\rangle\right) = \left(\bar{n}+\frac{1}{2}\right)\delta(t-t^\prime) ,\label{e23}
\end{align}
where $\bar{n}=[\exp(\hbar\omega_{m}/k_{B}T) -1]^{-1}$ is the mean number of the thermal excitations at the frequency of the mechanical mode,~$k_B$ is the Boltzmann constant and $T$ is the temperature of the environment.

The polaritons $\Psi$ and $\Phi$ might reasonably be called "bright" polaritons since they are damped with the rate $\gamma$. It is clear from Eq.~(\ref{e21}) that in the absence of the mechanical oscillator the system would consist of two completely decoupled bright polaritons. The effect of the mechanical oscillator that interests us most here is to introduce both shifts of the frequencies and coupling between the polaritons.

The exact treatment of the problem that involves quantum properties of the mechanical oscillator requires to deal with the system of nonlinear differential equations. The system of the equations is difficult to solve. Therefore, we use the linearization approach~\cite{bp03} by assuming that each operator of the system can be written as the sum of its steady-state mean value and a small fluctuation around the steady-state
\begin{align}
q &= q_{s} +\delta q ,\quad p = p_{s} +\delta p ,\nonumber\\
\Psi &= \Psi_{s} +\delta\Psi ,\quad \Phi = \Phi_{s} +\delta\Phi .\label{e24}
\end{align}
Note that this linearization approach is equivalent to the assumption of Gaussian distributions that describe fluctuations of the system around its stationary state. In this approach, Eq.~(\ref{e21}) decouples into a set of non-linear equations for the steady-state values and a set of differential equations for the fluctuation operators.

We first determine the average steady-state values of the operators. By taking the mean values of the operators, and then by setting the left-hand sides of Eq.~(\ref{e21}) to zero, we obtain the following steady-state solutions for the oscillator variables
\begin{eqnarray}
p_{s} = 0 ,\quad q_{s} = \frac{G_{0}}{\omega_{m}}\!\left|\Phi_{s}\cos\phi - \Psi_{s}\sin\phi\right|^{2}  ,\label{equps}
\end{eqnarray}
and the steady-state values of the polariton fields are found from the solution of two coupled nonlinear equations
\begin{align}
-E_{L}^{\ast}\sin\phi &=  \left[\gamma+i\left(\Delta_{q} -\tilde{\Omega}\right)\right]\Psi_{s} +\frac{1}{2}iGq_{s}\Phi_{s}  ,\nonumber\\
E_{L}^{\ast}\cos\phi &= \left[\gamma+i\left(\Delta_{q} +\tilde{\Omega}\right)\right]\Phi_{s} +\frac{1}{2}iGq_{s}\Psi_{s}  ,\label{e24a}
\end{align}
where $\Delta_{q} =\Delta_{L}-\frac{1}{2}q_{s}G_{0}$ and $\tilde{\Omega} =\Omega -\frac{1}{2}q_{s}G_{0}\cos(2\phi)$.

Under the linearization procedure and introduce the annihilation operator of the oscillating mirror, $\delta b = (\delta q +i\delta p)/\sqrt{2}$, the Langevin equations for the fluctuation operators satisfy the following set of differential equations
\begin{align}
\delta\dot{b} =& -\left(\frac{1}{2}\gamma_m +i\omega_{m}\!\right)\!\delta b +\frac{1}{2}\gamma_{m}\delta b^{\dag} -\frac{1}{2}iG_{\Psi}\!\left(\delta\Psi +\delta\Psi^{\dag} \right) \nonumber\\
&+\frac{1}{2}iG_{\Phi}\left(\delta\Phi +\delta\Phi^{\dag} \right) + \sqrt{\gamma_{m}}\xi ,\nonumber\\
\delta\dot{\Psi} =& -\left(\gamma+i\Delta_{\Psi}\right)\delta\Psi
       -\frac{1}{2}iG_{\Psi}\left(\delta b +\delta b^{\dag}\right) \nonumber\\
&-iG_{q}\delta\Phi + \sqrt{2\gamma}\Psi_{in} ,\nonumber\\
\delta\dot{\Phi} =& -\left(\gamma+i\Delta_{\Phi}\right)\delta\Phi
+\frac{1}{2}iG_{\Phi}\left(\delta b +\delta b^{\dag}\right) \nonumber\\
&-iG_{q}\delta\Psi +\sqrt{2 \gamma}\Phi_{in} .\label{e25w}
\end{align}
where we have chosen $\Delta_{\Psi} = (\Delta_{q} -\tilde{\Omega})$ and $\Delta_{\Phi} = (\Delta_{q} +\tilde{\Omega})$, and have introduced the abbreviations $G_{\Psi}\!=\!\sqrt{2}\!\left(\frac{1}{2}G\Phi_{s}\!-\!G_{s} \Psi_{s}\right)$, $G_{\Phi}\!=\!\sqrt{2}\!\left(G_{c} \Phi_{s}\!-\!\frac{1}{2}G\Psi_{s}\right)$ and $G_{q} =Gq_{s}/2$. Then we can get $G_{\Psi} =G_{\Phi}\tan\phi$. Also we see that~$G_{\Psi}$ and $G_{\Phi}$ are the effective coupling constants of the polaritons~$\Psi$ and $\Phi$ to the cavity field, respectively.

\section{Nano-mechanical entanglement}~\label{sec4}

We now apply Eqs.~(\ref{e25w}) explicitly to search for entanglement and correlations between the modes. Notice the presence of three different frequencies at which the fluctuation operators oscillate, $\omega_{m}, \Delta_{\Psi}$ and $\Delta_{\Phi}$. Thus, depending on whether we would like to entangle one or both polaritons to the oscillating mirror, we should choose $\omega_{m}$ to match to either the frequency of one of the polaritons or to the frequency $\Delta_{q}$, which is the mid-frequency of the two polaritons. To illustrate this, we introduce a rotating frame through the relations
\begin{align}
\delta{\tilde b} = \delta b\, {\rm e}^{i\omega_{m}t} ,\ \delta{\tilde\Psi} = \delta\Psi\, {\rm e}^{i\Delta_{\Psi} t} ,\ \delta{\tilde\Phi} = \delta\Phi\, {\rm e}^{i\Delta_{\Phi} t} ,
\end{align}
and find that in the rotating frame Eqs.~(\ref{e25w}) become
\begin{align}
\delta\dot{\tilde b} =& -\frac{1}{2}\gamma_m \delta{\tilde b} +\frac{1}{2}\gamma_{m}\delta {\tilde b}^{\dag}{\rm e}^{2i\omega_{m} t} + \sqrt{\gamma_{m}}\,\xi \,{\rm e}^{i\omega_{m} t}\nonumber\\
& -\frac{1}{2}iG_{\Psi}\!\left(\delta{\tilde\Psi}{\rm e}^{-i(\Delta_{\Psi}-\omega_{m}) t} +\delta{\tilde\Psi}^{\dag}{\rm e}^{i(\Delta_{\Psi} +\omega_{m}) t} \right) \nonumber\\
&+\frac{1}{2}iG_{\Phi}\!\left(\delta{\tilde\Phi}{\rm e}^{-i(\Delta_{\Phi}-\omega_{m}) t} +\delta{\tilde\Phi}^{\dag}{\rm e}^{i(\Delta_{\Phi}+\omega_{m}) t} \right)  ,\nonumber\\
\delta\dot{\tilde\Psi} =& -\gamma\delta{\tilde\Psi}
       -\frac{1}{2}iG_{\Psi}\left(\delta{\tilde b}{\rm e}^{i(\Delta_{\Psi}-\omega_{m}) t} +\delta{\tilde b}^{\dag}{\rm e}^{i(\Delta_{\Psi}+\omega_{m}) t}\right) \nonumber\\
&-iG_{q}\delta{\tilde\Phi}{\rm e}^{i(\Delta_{\Psi}-\Delta_{\Phi}) t}
+ \sqrt{2\gamma}\Psi_{in}{\rm e}^{i\Delta_{\Psi} t} ,\nonumber\\
\delta\dot{\tilde\Phi} =& -\gamma\delta{\tilde\Phi}
 +\frac{1}{2}iG_{\Phi}\left(\delta{\tilde b}{\rm e}^{i(\Delta_{\Phi}-\omega_{m}) t} +\delta{\tilde b}^{\dag}{\rm e}^{i(\Delta_{\Phi}+\omega_{m}) t}\right) \nonumber\\
&-iG_{q}\delta{\tilde\Psi}{\rm e}^{i(\Delta_{\Phi}-\Delta_{\Psi}) t}
+ \sqrt{2\gamma}\Phi_{in}{\rm e}^{i\Delta_{\Phi} t} .\label{e25p}
\end{align}
We see that the coupling terms of the polaritons $\delta{\tilde\Psi}$ and $\delta{\tilde\Phi}$ to the mirror operator oscillate in time with frequencies $\Delta_{\Psi}\pm\omega_{m}$ and $\Delta_{\Phi}\pm \omega_{m}$, respectively. When the equations are integrated over any measurable time interval, these oscillatory terms make a negligible contribution and can be ignored if they are different from zero. Therefore, in order for the coupling effects to be significant, we must have $\Delta_{\Psi} =\pm \omega_{m}$ or $\Delta_{\Phi} =\pm\omega_{m}$, when either of these terms become independent of time. These are optimal conditions for coupling of the mirror to either $\Psi$ or~$\Phi$ polariton.

An alternative choice of the new rotating frame
\begin{align}
\delta{\tilde b} = \delta b\, {\rm e}^{i\omega_{m}t} ,\ \delta{\tilde\Psi} = \delta\Psi\, {\rm e}^{i\Delta_{q} t} ,\ \delta{\tilde\Phi} = \delta\Phi\, {\rm e}^{i\Delta_{q} t} ,
\end{align}
results in the following transformed equations
\begin{align}
\delta\dot{\tilde b} =& -\frac{1}{2}\gamma_m \delta{\tilde b} +\frac{1}{2}\gamma_{m}\delta {\tilde b}^{\dag}{\rm e}^{2i\omega_{m} t} + \sqrt{\gamma_{m}}\,\xi \,{\rm e}^{i\omega_{m} t}\nonumber\\
&-\frac{1}{2}iG_{\Psi}\!\left(\delta{\tilde\Psi}{\rm e}^{-i(\Delta_{q}-\omega_{m}) t} +\delta{\tilde\Psi}^{\dag}{\rm e}^{i(\Delta_{q} +\omega_{m}) t} \right) \nonumber\\
&+\frac{1}{2}iG_{\Phi}\!\left(\delta{\tilde\Phi}{\rm e}^{-i(\Delta_{q}-\omega_{m}) t} +\delta{\tilde\Phi}^{\dag}{\rm e}^{i(\Delta_{q}+\omega_{m}) t} \right)  ,\nonumber\\
\delta\dot{\tilde\Psi} =& -\left(\gamma -i\tilde{\Omega}\right)\delta{\tilde\Psi} -iG_{q}\delta{\tilde\Phi} + \sqrt{2\gamma}\Psi_{in}{\rm e}^{i\Delta_{q} t} \nonumber\\
&-\frac{1}{2}iG_{\Psi}\left(\delta{\tilde b}{\rm e}^{i(\Delta_{q}-\omega_{m}) t} +\delta{\tilde b}^{\dag}{\rm e}^{i(\Delta_{q}+\omega_{m}) t}\right) ,\nonumber\\
\delta\dot{\tilde\Phi} =& -\left(\gamma +i\tilde{\Omega}\right)\delta{\tilde\Phi} -iG_{q}\delta{\tilde\Psi} + \sqrt{2\gamma}\Phi_{in}{\rm e}^{i\Delta_{q} t} \nonumber\\
&+\frac{1}{2}iG_{\Phi}\left(\delta{\tilde b}{\rm e}^{i(\Delta_{q}-\omega_{m}) t} +\delta{\tilde b}^{\dag}{\rm e}^{i(\Delta_{q}+\omega_{m}) t}\right)  .\label{e25q}
\end{align}
Now the exponential factors have frequency centered on $\Delta_{q}$, and in marked contrast to the previous situation, the choice of either $\Delta_{q} = \omega_{m}$ or $\Delta_{q}=- \omega_{m}$ would result in the simultaneous coupling of both polaritons to the oscillating mirror. It is interesting to note that the coupling between the polaritons is independent of the choice of the frequency $\omega_{m}$.

\subsection{Bipartite polariton-mirror coupling}

Let us first examine a bipartite coupling between the polariton $\Psi$ and the oscillating mirror. According to Eq.~(\ref{e25p}), this could be achieved by choosing $\Delta_{\Psi} =-\omega_{m}$, which has been shown as a necessary condition for entanglement between two bosonic modes~\cite{vt07}. In this case, the two modes are coupled by a parametric interaction~\cite{li05,li07}. The tuning of $\Delta_{\Psi} =\omega_{m}$ would not produce entanglement between the mirror and the polariton modes, since in this case the two modes are coupled by a linear-mixing interaction.

By choosing $\Delta_{\Psi}=-\omega_{m}$, after discarding the rapidly oscillating terms, we find from Eqs.~(\ref{e25p}) that the set of the differential equations~(\ref{e25p}) can be simplified to two separate sets of coupled differential equations for pairs $(\delta{\tilde b}^{\dag}, \delta{\tilde\Psi})$ and $(\delta{\tilde b}, \delta{\tilde\Psi}^{\dag})$. The equations of motion for the pair $(\delta{\tilde b}^{\dag}, \delta{\tilde\Psi})$ are
\begin{align}
\delta\dot{\tilde b}^{\dag} &= -\gamma \delta{\tilde b}^{\dag} +\frac{1}{2}iG_{\Psi}\delta{\tilde\Psi} + \sqrt{2\gamma}\,\tilde{\xi}^{\dag}(t) ,\nonumber\\
\delta\dot{\tilde\Psi} &= -\gamma\delta{\tilde\Psi} -\frac{1}{2}iG_{\Psi}\delta{\tilde b}^{\dag} + \sqrt{2\gamma}\tilde{\Psi}_{in}(t) ,\label{e25n}
\end{align}
where $\tilde{\xi}^{\dag}(t) =\xi^{\dag}\!\exp(\!-i\omega_{m}t)$, $\tilde{\Psi}_{in}(t)=\Psi_{in}\!\exp(\!-i\omega_{m}t)$, and we have put $\gamma_{m}=2\gamma$. Note that the two coupled modes oscillate at the same frequencies. This may result in the so-called {\it one-colour} entanglement, i.e. entanglement between two modes of the same frequency.

\subsection{Identification of entanglement from the squeezing condition}

In order to show that entanglement and squeezing can be created between the $\Psi$ and $b$ modes, we introduce in-phase and out-of-phase quadrature components of the fluctuation operators
\begin{align}
\delta\tilde{\Psi}_{x} &= \frac{1}{\sqrt{2}}\!\left(\delta\tilde{\Psi}{\rm e}^{-i\psi}+\delta\tilde{\Psi}^\dag{\rm e}^{i\psi}\right) ,\nonumber\\
\delta\tilde{\Psi}_{y} &= \frac{1}{\sqrt{2}i}\!\left(\delta\tilde{\Psi}{\rm e}^{-i\psi}-\delta\tilde{\Psi}^\dag{\rm e}^{i\psi}\right) ,\nonumber\\
\delta\tilde{\Phi}_{x} &= \frac{1}{\sqrt{2}}\!\left(\delta\tilde{\Phi}{\rm e}^{-i\psi} +\delta\tilde{\Phi}^\dag{\rm e}^{i\psi}\right) ,\nonumber\\
\delta\tilde{\Phi}_{y} &= \frac{1}{\sqrt{2}i}\!\left(\delta\tilde{\Phi}{\rm e}^{-i\psi} -\delta\tilde{\Phi}^\dag{\rm e}^{i\psi}\right) ,\label{e26p}
\end{align}
and
\begin{align}
\delta\Lambda_{x} &= \frac{1}{\sqrt{2}}\left(\delta\tilde{\Psi}_x -\delta q\right) ,\quad
\delta\Upsilon_{x} = \frac{1}{\sqrt{2}}\left(\delta\tilde{\Psi_x} +\delta q\right) ,\nonumber\\
\delta\Lambda_{y} &= \frac{1}{\sqrt{2}i}\left(\delta\tilde{\Psi}_y -\delta p\right) ,\quad
\delta\Upsilon_{y} = \frac{1}{\sqrt{2}i}\left(\delta\tilde{\Psi}_y +\delta q\right) ,
\end{align}

In order to see if there are correlations existing between the polariton $\Psi$ and the mechanical mode $b$, we must examine properties of the sum operators, either $\delta\Upsilon_{x}$ or $\delta\Upsilon_{y}$ and the difference operators, either $\delta\Lambda_{x}$ or $\delta\Lambda_{y}$ that act on both systems. The variances in the sum and difference operators are given by
\begin{align}
&\Delta\left(\delta\Upsilon_{x}\right)^{2} = \langle\delta\Upsilon_{x}^{2}\rangle -\langle\delta\Upsilon_{x}\rangle^{2} \nonumber\\
&= \frac{1}{2}\left[\Delta(\delta\tilde{\Psi}_{x})^{2} +\Delta(\delta q)^{2} + 2\left(\langle\delta\tilde{\Psi}_{x}\delta q\rangle -\langle\delta\tilde{\Psi}_{x}\rangle\langle\delta q\rangle\right)\right] ,\nonumber\\
&\Delta\left(\delta\Lambda_{x}\right)^{2} = \langle\delta\Lambda_{x}^{2}\rangle -\langle\delta\Lambda_{x}\rangle^{2} \nonumber\\
&= \frac{1}{2}\left[\Delta(\delta\tilde{\Psi}_{x})^{2} +\Delta(\delta q)^{2} - 2\left(\langle\delta\tilde{\Psi}_{x}\delta q\rangle -\langle\delta\tilde{\Psi}_{x}\rangle\langle\delta q\rangle\right)\right]  .\label{e55}
\end{align}
The product of these two variances is
\begin{align}
\Delta\left(\delta\Upsilon_{x}\right)^{2}\Delta\left(\delta\Lambda_{x}\right)^{2} &=
\frac{1}{4}\left[\Delta(\delta\tilde{\Psi}_{x})^{2} +\Delta(\delta q)^{2}\right]^{2}\nonumber\\
&- \left(\langle\delta\tilde{\Psi}_{x}\delta q\rangle -\langle\delta\tilde{\Psi}_{x}\rangle\langle\delta q\rangle\right)^{2}  .\label{e56}
\end{align}
Hence, we have an uncertainty relation
\begin{align}
\sqrt{\Delta\left(\delta\Upsilon_{x}\right)^{2}\Delta\left(\delta\Lambda_{x}\right)^{2}} \leq
\frac{1}{2}\left[\Delta(\delta{\tilde\Psi}_{x})^{2} +\Delta(\delta q)^{2}\right] ,\label{e57}
\end{align}
with equality only holding if the systems are uncorrelated.

Solving Eq.~(\ref{e25n}) for the steady-state $(t\rightarrow\infty)$, we find
\begin{align}
\Delta\left(\delta\Upsilon_{x}\right)^{2} &= \frac{\gamma (\bar{n}+1)}{2\left(\gamma -\frac{1}{2}G_{\Psi}\right)} ,\nonumber\\
\Delta\left(\delta\Lambda_{x}\right)^{2} &= \frac{\gamma (\bar{n}+1)}{2\left(\gamma +\frac{1}{2}G_{\Psi}\right)} ,\label{e58}
\end{align}
and
\begin{align}
\Delta\left(\delta{\tilde\Psi}_{x}\right)^{2} &= \frac{1}{2}\left[\frac{\gamma^{2}(\bar{n}+1)}{\left(\gamma^{2} -\frac{1}{4}G_{\Psi}^{2}\right)} -\bar{n}\right] ,\nonumber\\
\Delta\left(\delta q\right)^{2} &= \frac{1}{2}\left[\frac{\gamma^{2}(\bar{n}+1)}{\left(\gamma^{2} - \frac{1}{4}G_{\Psi}^{2}\right)} +\bar{n}\right] .\label{e59}
\end{align}
It is seen that the variances $\Delta(\delta{\tilde\Psi}_{x})^{2}$ and $\Delta\left(\delta q\right)^{2}$ are both larger than their vacuum limits, i.e. $1/2$. Thus, both modes are not themselves squeezed and display thermal fluctuations. However, substituting Eqs.~(\ref{e58}) and (\ref{e59}) into Eq.~(\ref{e57}), we find that the polariton and the mechanical modes are correlated when
\begin{align}
\frac{\gamma^{2}}{\left(\gamma^{2} -\frac{1}{4}G_{\Psi}^{2}\right)} > 1 ,
\end{align}
which is always satisfied as long as $G_{\Psi}\neq 0$. We stress that the above inequality is necessary but not sufficient condition for squeezing (entanglement) between the modes. In other words, the modes could be correlated but not enough to beat the quantum limit for fluctuations. Equivalently, we may say that the modes are correlated classically and the quantum limit can be beaten only if the modes exhibit quantum correlations.

The sufficient condition for squeezing is that the variance $\Delta\left(\delta\Upsilon_{x}\right)^{2}$ is reduced below the limit for quantum fluctuations, i.e. below $1/2$. It is easily verified from Eq.~(\ref{e58}) that the correlations between the modes will lead to squeezing in the superposition $\delta\Upsilon_{x}$ when $G_{\Psi}>2\bar{n}\gamma$.
However, there is an upper limit on $G_{\Psi}$ imposed by the condition of stable steady-state solutions of Eqs.~(\ref{e25n}). One can easily find from Eqs.~(\ref{e25n}) that the stability of the steady-state solutions requires $G_{\Psi}<2\gamma$. Thus, combining the stability and squeezing conditions, we find that the necessary and sufficient condition for entanglement between the polariton $\Psi$ and the vibrating mirror mode~is
\begin{align}
2\bar{n}\gamma < G_{\Psi} < 2\gamma ,\label{sqc}
\end{align}
which, on the other hand, indicates that the modes can be entangled only if $\bar{n}<1$.

Since the modes are correlated for any $G_{\Psi} < 2\gamma$ and the condition for squeezing (entanglement) is that $G_{\Psi}$ must be greater than $2\bar{n}\gamma$, there is evidently a significant restriction on the strength of the coupling of the vibrating mirror to the polariton mode. The condition (\ref{sqc}) for entanglement is essentially similar to that of a microcavity mode and a vibrating mirror treated by Vitali {\it et al.}~\cite{vt07}.

\subsection{Violation of the Cauchy-Schwartz inequality and anomalous correlations}

An alternative and in fact more elegant way to study entanglement between two modes $(A,B)$ is the Cauchy-Schwartz inequality~\cite{mw95}
\begin{align}
\chi_{(A,B)} = \frac{g^{(2)}_{A}g^{(2)}_{B}}{\left(g^{(2)}_{AB}\right)^{2}} > 1 .\label{eq62}
\end{align}
Here, $\chi_{(A,B)}$ is the so-called Cauchy-Schwartz parameter,
\begin{align}
g^{(2)}_{AB} = \frac{\langle A^{\dagger}B^{\dagger}AB\rangle}{\langle A^{\dagger}A\rangle\langle B^{\dagger}B\rangle}
\end{align}
is the normalized second-order cross correlation function, and
\begin{align}
g^{(2)}_{A} = \frac{\langle A^{\dagger 2}A^{2}\rangle}{\langle A^{\dagger}A\rangle^{2}} ,\quad
g^{(2)}_{B} = \frac{\langle B^{\dagger 2}B^{2}\rangle}{\langle B^{\dagger}B\rangle^{2}} ,
\end{align}
are the normalized intensity autocorrelation functions of the modes $A$ and $B$, respectively.

Since the modes obey the Gaussian statistics, the correlation functions can be readily related to coherence functions
\begin{align}
g^{(2)}_{n} &= 2 + \left|\eta_{(n,n)}\right|^{2}  ,\quad n=A,B, \nonumber\\
g^{(2)}_{AB} &= 1 + \left|\gamma_{(A,B)}\right|^{2} +\left|\eta_{(A,B)}\right|^{2} ,\label{eq38}
\end{align}
where $\left|\gamma_{(A,B)}\right|$, defined as
\begin{align}
\left|\gamma_{(A,B)}\right| = \frac{|\langle A^{\dag}B\rangle|}{\sqrt{\langle A^{\dag}A\rangle\langle B^{\dag}B\rangle}} ,\label{eq32}
\end{align}
is the degree of the first-order coherence,
\begin{align}
\left|\eta_{(A,A)}\right| = \frac{\left|\langle A^{2}\rangle\right|}{\langle A^{\dag}A\rangle} ,\quad \left|\eta_{(B,B)}\right| = \frac{\left|\langle B^{2}\rangle\right|}{\langle B^{\dag}B\rangle} ,
\end{align}
are degrees of the so-called "anomalous" autocorrelation, and
\begin{align}
\left|\eta_{(A,B)}\right| &= \frac{\left|\langle AB\rangle\right|}{\sqrt{\langle A^{\dag}A\rangle\langle B^{\dag}B\rangle}} ,\label{eq39}
\end{align}
is the degree of the anomalous cross correlation~\cite{ks80,lo84,ag86,hr87,ft88,sl11}.
Equation~(\ref{eq38}) shows that the second-order autocorrelation functions depend on the anomalous autocorrelation whereas the second-order cross correlation function depends on the first-order coherence and the anomalous cross correlation.

We now determine conditions under which the correlation between the polariton $\Psi$ and the mechanical mode~$b$ violate the Cauchy-Schwartz inequality (\ref{eq62}).
Following the method introduced in the book~\cite{bk30}, we can easily solve Eqs.~(\ref{e25n}) to find that in the steady-state
\begin{align}
\langle\delta\Psi^{2}\rangle = \langle\delta b^{2}\rangle = 0 ,\quad \langle\delta\Psi^{\dagger}\delta b\rangle = 0 ,
\end{align}
and
\begin{align}
\langle\delta\Psi\delta b\rangle =-\frac{i}{4}\left(\bar{n}+1\right)\frac{\gamma G_{\Psi}}{\gamma^{2} -\frac{1}{4}G_{\Psi}^{2}} ,
\end{align}
which together with
\begin{align}
\langle\delta\Psi^{\dagger}\delta\Psi\rangle &= \frac{1}{8}\left(\bar{n}+1\right)\frac{G_{\Psi}^{2}}{\gamma^{2} -\frac{1}{4}G^{2}_{\Psi}} ,\nonumber\\
\langle\delta b^{\dagger}\delta b\rangle &= \frac{1}{2}\left[\left(\bar{n}-1\right) +\frac{\left(\bar{n}+1\right)\gamma^{2}}{\gamma^{2} -\frac{1}{4}G_{\Psi}^{2}}\right] ,
\end{align}
give the following condition for the violation of the Cauchy-Schwartz inequality
\begin{align}
\left|\eta_{(\Psi,b)}\right|^{2} = \frac{\left(\bar{n}+1\right)\gamma^{2}}{\left[2\bar{n}\gamma^{2} -\frac{1}{4}\left(\bar{n}-1\right)G_{\Psi}^{2}\right]} > 1 .\label{eq70}
\end{align}
We emphasize that the inequality (\ref{eq70}) is the necessary and sufficient condition for entanglement between the modes. It is easily verified that the inequality is satisfied as long as $\bar{n}<1$ and $G_{\Psi}< 2\gamma$. This shows that the Cauchy-Schwartz inequality is violated under the same conditions the squeezing is generated between the modes, see Eq.~(\ref{sqc}). Moreover, the inequality (\ref{eq70}) implies that the violation of the Cauchy-Schwartz inequality is achieved simply by the requirement that the anomalous cross coherence is greater than the product of the intensities of the modes.
It is interesting to note that the condition (\ref{eq70}) corresponds to the case of each of the modes being in the thermal state, $g^{(2)}_{\Psi}=g^{(2)}_{b}=2$, with no the first-order coherence between them, $\left|\gamma_{(\Psi,b)}\right| = 0$. Thus, the fields of the polariton and mechanical modes are entangled but there is no interference.

We now illustrate the above predicted limitations for the occurrence of entanglement and also provide the quantitative value of the bipartite entanglement. For this, we use the logarithmic negativity that is known as the necessary and sufficient condition for entanglement of two-mode Gaussian states~\cite{si00,as04}
\begin{eqnarray}
E_{N}  = {\rm max}\left\{0,-\log_{2}\left[2V_{s}\right]\right\} ,\label{e58e}
\end{eqnarray}
where $V_{s}$ is the smallest sympletic eigenvalue of the partially transposed correlation (covariance) matrix ${\bf V}$, with elements
\begin{align}
V_{ij}=\langle u_i(\infty)u_j(\infty) + u_j(\infty)u_i(\infty)\rangle/2 ,\label{e59e}
\end{align}
where $u_i(\infty)$ is the steady-state value of the $i$th component of the vector $\vec{u}$:
\begin{align}
\vec{u} =\left(\delta\tilde{q},\delta\tilde{p},\delta\tilde{\Psi}_{x},\delta\tilde{\Psi}_{y}\right)^T ,\label{e60e}
\end{align}
with $\delta\tilde{q}=(\delta\tilde{b}+\delta\tilde{b}^{\dagger})/\sqrt{2}, \delta\tilde{p}=(\delta\tilde{b}-\delta\tilde{b}^{\dagger})/\sqrt{2}i$, and $\delta\tilde{\Psi}_{x},\delta\tilde{\Psi}_{y}$ defined in Eq.~(\ref{e26p}).

The steady-state values are readily calculated using the equations of motion (\ref{e25n}), from which we get the following matrix equation
\begin{align}
\dot{\vec{u}}(t) = {\bf A} \vec{u}(t) +\sqrt{2 \gamma}\,\vec{\eta}(t) ,\label{e61e}
\end{align}
where the drift matrix ${\bf A}$ is given by
\begin{align}
\bf{A}& = \left(\begin{array}{cccc}
-\gamma&0&0&-\frac{1}{2}G_{\Psi}\\
0&-\gamma&-\frac{1}{2}G_{\Psi}&0\\
0&-\frac{1}{2}G_{\Psi}&-\gamma&0\\
-\frac{1}{2}G_{\Psi}&0&0&-\gamma
\end{array}\right) ,\label{e62e}
\end{align}
and
\begin{align}
\vec{\eta}(t) = \left(\tilde{q}_{in}(t),\tilde{p}_{in}(t),\tilde{\Psi}_{in}^{x}(t),\tilde{\Psi}_{in}^{y}(t)\right)^{T} .\label{e63e}
\end{align}
The matrix equation (\ref{e61e}) is a simple first order differential equation with time-independent coefficients, and is solved by a direct integration. The formal solution is given by
\begin{align}
\vec{u}(t) = \vec{u}(0){\rm e}^{At} +\sqrt{2 \gamma} \int_{0}^{t} dt^{\prime} \, \vec{\eta}(t-t^{\prime}){\rm e}^{At^{\prime}} ,\label{e63b}
\end{align}
where $\vec{u}(0)$ is the vector of initial values of the components. For the steady-state, we take the limit of Eq.~(\ref{e63b}) as $t\rightarrow\infty$.

Since the noise $\xi(t)$ is $\delta$-correlated, so that it describes a Markovian process, the steady-state correlation matrix is then derived from the following equation~\cite{bp03}:
\begin{align}
{\bf AV}+{\bf VA}^T = -{\bf D} ,\label{e64e}
\end{align}
where  ${\bf D} =\textrm{diag}[(2\bar{n}+1)\gamma,(2\bar{n}+1)\gamma,\gamma,\gamma]$ is the diffusion matrix stemming from the noise correlations.

\begin{figure}[h]
\begin{center}
\begin{tabular}{c}
\includegraphics[width=0.95\columnwidth]{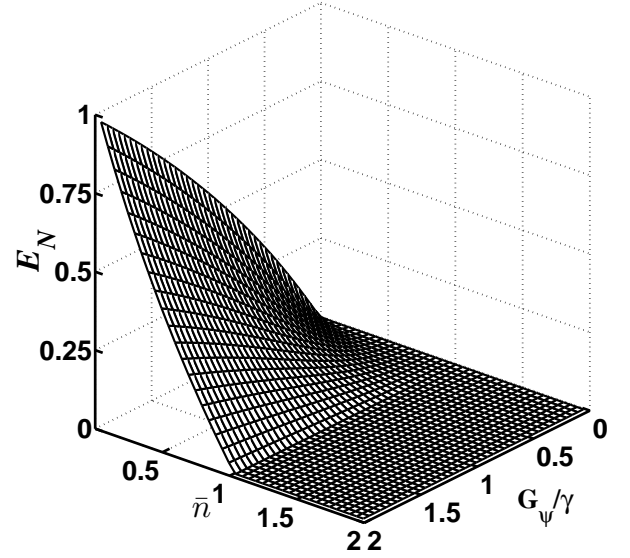}
\end{tabular}
\end{center}
\caption[nsa]{ \label{mirfig2} Variation of the logarithmic negativity $E_{N}$ with $\bar{n}$ and~$G_{\Psi}/\gamma$.}
\end{figure}

Figure~\ref{mirfig2} illustrates the dependence of the logarithmic negativity on $\bar{n}$ and~$G_{\Psi}$. It is apparent that the modes are entangled for $\bar{n}<1$ and $G_{\Psi}<2\gamma$. Once $\bar{n}$ is greater than 1, entanglement becomes impossible irrespective of $G_{\Psi}$.

The dependence of the second-order correlation functions~(\ref{eq38}) on different kind of coherence functions allows us to determine which of the coherences work for and which work against the creation of a strong entanglement between the modes. A simple analysis of Eqs.~(\ref{eq62}) and (\ref{eq38}) shows that in general the Cauchy-Schwartz inequality is violated when
\begin{align}
(2\!+\!|\eta_{(A,A)}|^{2})(2\!+\!|\eta_{(B,B)}|^{2}) < \left(1\!+\!|\gamma_{(A,B)}|^{2}\!+\!|\eta_{(A,B)}|^{2}\right)^{2} .\label{eq71}
\end{align}
It follows from Eq.~(\ref{eq71}) that the general condition for the violation of the Cauchy-Schwartz inequality cannot be viewed as exclusively dependent on $|\eta_{(A,B)}|^{2}$. One could notice that the Cauchy-Schwartz inequality would be easier to violate if the first and second order cross correlations were simultaneously created while the anomalous autocorrelations were kept zero. However, we shall demonstrate below that this situation is unfounded, that the simultaneous creation of the cross correlation functions is equally effective in creating the anomalous autocorrelations.

\subsection{Squeezed vacuum environment}

When evaluating the correlation function $\langle \delta\Psi^{\dagger}\delta b\rangle$, one can find that the function  depends solely on the noise two-photon correlation functions $\langle \Phi^\dag_{in}(t) \Phi^\dag_{in}(t^\prime)\rangle$ and $\langle\tilde{\xi}^{\dagger}(t)\tilde{\xi}^{\dagger}(t^{\prime}\rangle$. Therefore, the correlation function $\langle \delta\Psi^{\dagger}\delta b\rangle$ could be different from zero when either the polariton or the mirror were located in the environment whose modes exhibit nonzero two-photon correlations. An example of such environment is a squeezed vacuum field which in the case of the mirror environment is determined by Eq.~(\ref{e23}) and the following second moments
\begin{align}
&\langle\tilde{\xi}(t)\tilde{\xi}^{\dagger}(t^{\prime})\rangle = \left(\bar{n}\!+\!1\right)\delta(t-t^{\prime}) ,\ \langle\tilde{\xi}^{\dagger}(t)\tilde{\xi}(t^{\prime})\rangle = \bar{n}\delta(t-t^{\prime}) ,\nonumber\\
&\langle\tilde{\xi}(t)\tilde{\xi}^(t^{\prime}\rangle = m\delta(t+t^{\prime}) ,\ \langle\tilde{\xi}^{\dagger}(t)\tilde{\xi}^{\dagger}(t^{\prime}\rangle = m^{\ast}\delta(t+t^{\prime}) ,\label{eq72a}
\end{align}
where $\bar{n}$ is the squeezing photon number and $|m|\leq \sqrt{\bar{n}(\bar{n}+1)}$ measures the strength of two-photon correlations~\cite{df04}. Thus, if the mirror were oscillating in the squeezed vacuum field, this would create the first-order coherence between the modes.

Let us apply these considerations explicitly to the first-order correlation function $\langle \delta\Psi^{\dagger}\delta b\rangle$. If the mirror oscillates in the squeezed vacuum field, we readily find that
\begin{eqnarray}
\langle \delta\Psi^{\dagger}\delta b\rangle = \frac{i\gamma m G_{\Psi}}{4\left(\gamma^{2}-\frac{1}{4}G^{2}_{\Psi}\right)} ,\label{eq73}
\end{eqnarray}
which is nonzero as long as the polariton and the mirror are coupled to each other. When Eq.~(\ref{eq73}) is substituted into Eq.~(\ref{eq32}), we find for the degree of the first-order coherence
\begin{align}
\left|\gamma_{(\Psi,b)}\right| = \frac{\gamma|m|}{\left\{\left(\bar{n}+1\right)\left[2\bar{n}\gamma^{2}
-\frac{1}{4}\left(\bar{n}-1\right)G_{\Psi}^{2}\right]\right\}^{1/2}} ,\label{eq74}
\end{align}
which is less than 1 in general, and becomes unity only in the limit of $G_{\Psi}=2\gamma$ and $\bar{n}\gg 1$.
Expression (\ref{eq74}) shows that the interaction of the oscillating mirror with the squeezed vacuum field results in the first-order coherence between the polariton and the mechanical modes. That happens because the two-photon correlations present in the squeezed field have the effect of inducing stimulated two-photon processes which, it turns out, are sufficient for the polariton and mirror fields to become mutually coherent. Equivalently, the oscillating mirror that scatters photons from the squeezed vacuum to the polariton mode gives rise to phase locking between the polariton and mechanical modes. The correlations in the squeezed vacuum field have therefore induced coherence between the polariton and mechanical modes. It is interesting to note that the squeezed correlations do not effect the anomalous cross correlation $\langle\delta\Psi\delta b\rangle$.

The creation of the first-order coherence should, according to Eq.~(\ref{eq71}), enhance the violation of the Cauchy-Schwartz inequality. However, this is not the case, because the squeezed vacuum not only creates the first-order coherence but also the anomalous autocorrelations. It is easy to find that
\begin{align}
\langle\delta\Psi^{2}\rangle &= -\frac{1}{8}m \frac{G^{2}_{\Psi}}{\left(\gamma^{2} -\frac{1}{4}G^{2}_{\Psi}\right)} ,\nonumber\\
\langle\delta b^{2}\rangle &= \frac{1}{2}m\left[1 +\frac{\gamma^{2}}{\left(\gamma^{2}-\frac{1}{4}G^{2}_{\Psi}\right)}\right] ,
\end{align}
which shows that in both modes the anomalous autocorrelations are induced by the squeezed vacuum. The degrees of the anomalous coherences are then given by
\begin{align}
\left|\eta_{(\Psi,\Psi)}\right| &= \frac{|\langle\delta\Psi^{2}\rangle|}{\langle\delta\Psi^{\dagger}\delta\Psi\rangle} = \frac{|m|}{\bar{n}+1} ,\nonumber\\
\left|\eta_{(b,b)}\right| &= \frac{|\langle\delta b^{2}\rangle|}{\langle\delta b^{\dagger}\delta b\rangle} = \frac{|m|\left(2\gamma^{2} -\frac{1}{4}G^{2}_{\Psi}\right)}{\left[2\bar{n}\gamma^{2} -\frac{1}{4}\left(\bar{n}-1\right)G^{2}_{\Psi}\right]} .
\end{align}
Clearly the squeezed vacuum conspires to create the anomalous autocorrelation functions. Note that $|\gamma_{(\Psi,b)}|\leq \sqrt{|\eta_{(\Psi,\Psi)}||\eta_{(b,b)}|}$ and the equality holds at the threshold value of $G_{\Psi}=2\gamma$. It can also be seen that the anomalous autocorrelation of the mechanical mode is greater than that of the polariton mode and the equality between the autocorrelations is achieved when~$G_{\Psi}=2\gamma$.

The presence of the anomalous autocorrelations alters the condition (\ref{eq71}) for the violation of the Cauchy-Schwartz inequality. It is clear that the left-hand-side of Eq.~(\ref{eq71}) is enhanced by the anomalous autocorrelation functions. Thus the inequality (\ref{eq71}) can be harder to achieve. Needless to say, the creation of the first-order coherence among the modes is achieved at the cost of a corresponding decrease in the entanglement between the modes.
\begin{figure}[h]
\begin{center}
\begin{tabular}{c}
\includegraphics[width=0.9\columnwidth]{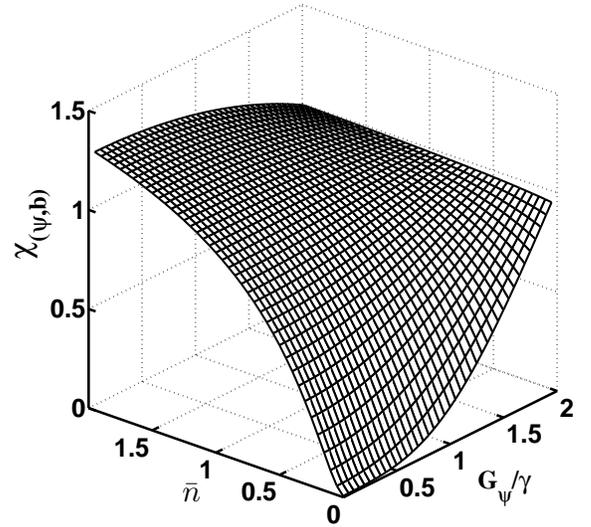}
\end{tabular}
\end{center}
\caption[nsa]{ \label{mirfig3} Variation of the Cauchy-Schwartz parameter $\chi_{(\Psi,b)}$ with $\bar{n}$ and~$G_{\Psi}/\gamma$ for $|m|=\sqrt{\bar{n}(\bar{n}+1)}$.}
\end{figure}

Figure~\ref{mirfig3} illustrates the Cauchy-Schwartz parameter as a function of $\bar{n}$ and $G_{\Psi}$ for the maximally correlated squeezed vacuum field, $|m|=\sqrt{\bar{n}(\bar{n}+1)}$. The effect of including the squeezing correlations is clearly to restricts the range of $\bar{n}$ and~$G_{\Psi}$ over which the modes are entangled.

We may conclude this section that for the best conditions to entangle two degenerate modes is a situation of mutually incoherent modes each being in the thermal state.

\section{Three-mode coherence and entanglement}~\label{sec5}

In this section we determine conditions for correlations and entanglement when both polaritons $\Psi$ and $\Phi$ are simultaneously coupled to the mechanical mode. This is a situation of the three-mode interaction and appears in the system when the frequency of the mechanical mode~$\omega_{m}$ is tuned to the midd-frequency of the two polaritons, i.e. when~$\omega_{m}=-\Delta_{q}$. We also briefly study the two-colour entanglement that may occur between two modes of different frequencies.

When we chose $\Delta_{q}=-\omega_{m}$ and make the secular approximation, we find that Eqs.~(\ref{e25q}) simplify to two separate sets of three coupled differential equations for $(\delta{\tilde b}^{\dag}, \delta{\tilde\Psi}, \delta{\tilde\Phi})$ and $(\delta{\tilde b}, \delta{\tilde\Psi}^{\dag}, \delta{\tilde\Phi}^{\dag})$. For example, the equations of motion for the set $(\delta{\tilde b}^{\dag}, \delta{\tilde\Psi}, \delta{\tilde\Phi})$ are
\begin{align}
\delta\dot{\tilde b}^{\dag} &= -\frac{1}{2}\gamma_m \delta{\tilde b}^{\dag} +\frac{1}{2}iG_{\Psi}\delta{\tilde\Psi} -\frac{1}{2}iG_{\Phi}\delta{\tilde\Phi} + \sqrt{\gamma_{m}}\,\tilde{\xi}^{\dag}(t)  ,\nonumber\\
\delta\dot{\tilde\Psi} &= -\left(\gamma -i\tilde{\Omega}\right)\delta{\tilde\Psi} -iG_{q}\delta{\tilde\Phi} -\frac{1}{2}iG_{\Psi}\delta{\tilde b}^{\dag} + \sqrt{2\gamma}\tilde{\Psi}_{in}(t)  ,\nonumber\\
\delta\dot{\tilde\Phi} &= -\left(\gamma +i\tilde{\Omega}\right)\delta{\tilde\Phi} -iG_{q}\delta{\tilde\Psi} +\frac{1}{2}iG_{\Phi}\delta{\tilde b}^{\dag} + \sqrt{2\gamma}\tilde{\Phi}_{in}(t) ,\label{e55g}
\end{align}
where $\tilde{\Phi}_{in}(t) = \Phi_{in}\exp(-i\omega_{m} t)$. Equations (\ref{e55g}) are quite different from Eq.~(\ref{e25n}) that now both polaritons are coupled to the mechanical mode and the coupling is of the type of a parametric interaction. Normally, we would expect that this kind of coupling should result in an entanglement between the polaritons. As we shall see below, this hope is unfounded, the coupling of the polaritons with the mechanical mode by the parametric interaction results in the first-order rather than a second-order coherence between the polaritons.

Before going into detailed studies of the conditions for entanglement between the modes, we first comment about certain general features of the simultaneous coupling of the polaritons to the mechanical mode that follow from Eq.~(\ref{e55g}). We see that the effect of the mechanical mode on the dynamics of the polaritons is twofold. Firstly, the mechanical mode couples the polaritons to each other with the coupling strength~$G_{q}$. This indicates that in the presence of the oscillating mirror, the polaritons $\Psi$ and $\Phi$ are no longer the eigenstates of the system. Secondly, the mirror couples to the polaritons with different coupling strengths $G_{\Psi}$ and $G_{\Phi}$. The magnitude of the coupling strengths depends on whether a given polariton is maximally entangled or not. When the polaritons are maximally entangled $G_{\Psi}=G_{\Phi}$, otherwise $G_{\Psi}\neq G_{\Phi}$ when the polaritons are non-maximally entangled.

The presence of the coupling between the polaritons prompts us to make an unitary transformation to obtain 'new' orthogonal polariton modes. It is easily shown that the annihilation operators of the orthogonal superposition modes are of the form
\begin{align}
\delta\Theta &= \cos(\phi +\varphi)\delta C_{1} -\sin(\phi +\varphi)\delta a ,\nonumber\\
\delta\Pi &= \sin(\phi +\varphi)\delta C_{1} + \cos(\phi +\varphi)\delta a  .\label{e59m}
\end{align}
where the angle $\varphi$ is defined by
\begin{align}
\cos^{2}\varphi = \frac{1}{2} +\frac{\tilde{\Omega}}{2U} ,
\end{align}
with $U = \sqrt{\tilde{\Omega}^{2} + G_{q}^{2}}$. Note that the angle $\varphi$ belongs to the interval $[0,\pi/4]$.

Several interesting features can be found in Eq.~(\ref{e59m}). Firstly, we note that the initially maximally entangled exciton and the cavity fields, i.e. $\phi =\pi/4$ and $G_{q}=0$, become non-maximally entangled when the mechanical mode is included, $G_{q}\neq 0$. Thus, the effect of the mechanical mode on the initial maximally entangled exciton and the cavity  modes is to destroy the superposition (polariton) modes. Secondly, if initially the exciton and the cavity fields were disentangled, i.e. $\phi = 0$, they remain disentangled even in the presence of the mechanical mode because in this case $G_{q}=0$ and then $\varphi =0$. Thirdly, an initial non-maximally entangled state between the exciton and the cavity modes, $0<\phi<\pi/2$ can be transferred to the maximally entangled state by the mechanical effect. It happens when $\phi +\varphi =\pi/4$. However, the maximally entangled state can be reached only for initial superpositions with $\phi<\pi/4$. Otherwise, for $\phi>\pi/4$, the superposition cannot be transferred by the mechanical effect into the maximally entangled state.

We now turn to the calculation of the correlation functions between the modes which provide the information about coherence and entanglement between the modes. In terms of the transformed operators $\delta\Theta$ and $\delta\Pi$, the equations of motion Eqs.~(\ref{e55g}) take the form
\begin{align}
\left(
\begin{array}{c}
\delta\dot{\tilde b}^{\dag}\\
\delta\dot{\Theta}\\
\delta\dot{\Pi}
\end{array}
\right) &= -\left(
\begin{array}{ccc}
\gamma & -\frac{1}{2}iG_{\theta} & -\frac{1}{2}iG_{\pi} \\
\frac{1}{2}iG_{\theta} & \left(\gamma +iU\right)&0\\
\frac{1}{2}iG_{\pi} & 0&\left(\gamma -iU\right)\\
\end{array}
\right) \left(
\begin{array}{c}
\delta{\tilde b}^{\dag}\\
\delta\Theta\\
\delta\Pi
\end{array}
\right) \nonumber\\
&+\sqrt{2\gamma}\left(
\begin{array}{c}
\tilde{\xi}^{\dag}(t)\\
\Theta_{in}(t)\\
\Pi_{in}(t)
\end{array}
\right)
,\label{e55b}
\end{align}
where $G_{\theta}= (G_{\Psi}\cos\varphi +G_{\Phi}\sin\varphi)$ and $G_{\pi}= (G_{\Psi}\sin\varphi -G_{\Phi}\cos\varphi)$ are effective couplings of the transformed polariton modes to the cavity mode. It is clear from Eq.~(\ref{e55b}) that the transformed polaritons are independent of each other, they oscillate with different frequencies, shifted from $\omega_{m}$ by an amount $\pm U$, and both are coupled to the mechanical mode by the parametric amplification process. Notice that $G_{\theta}=-G_{\pi}\tan(\phi +\varphi)$.

\subsection{Two-colour entanglement}

Since the modes oscillate at different frequencies, we may apply the equations (\ref{e55b}) to the problem of two-colour entanglement~\cite{vc05,cb10,wg10,lg10,sl10,co11,tan11}. If we assume that $G_{\pi}=0$, we then can ignore coupling of the $\delta\Pi$ polariton to the mechanical mode and limit ourselves to considering the case of two coupled modes of different frequencies. This could takes place, for example, when the cavity field frequency were on resonance with the exciton frequency $(\delta=0)$ and the coupling of the mirror to the cavity mode were much stronger than the coupling of the atoms to the cavity mode, i.e. $G_{q}\gg\tilde{\Omega}$. In this case, $\phi=\varphi =\pi/4$ and then it is easy to find that~$G_{\pi}=0$.

To check the conditions for the stable steady-state solutions, we put $G_{\pi}=0$ in the $3\times 3$ matrix appearing in Eq.~(\ref{e55b}), and find the following eigenvalues
\begin{align}
\lambda_{1} =\gamma-iU ,\quad \lambda_{2,3} = \left(\gamma \pm \sqrt{G_{\theta}^{2} -U^{2}}\right) +\frac{1}{2}iU .
\end{align}
We see that a threshold occurs for the coupling strength at $G_{\theta}=U$ where the eigenvalues $\lambda_{2,3}$ change character. Below the threshold, $G_{\theta}<U$, the real parts of the eigenvalues are positive irrespective of the values of the parameters involved. Above the threshold, $G_{\theta}>U$, the real parts are positive when $\sqrt{G_{\theta}^{2} -U^{2}}<\gamma$. We should stress here that under the condition of $G_{q}\gg\tilde{\Omega}$, taken here, the inequality $G_{\theta}<U$, i.e. the below threshold situation always holds. Therefore, as long as $G_{\theta}<U$ there are no restrictions on the parameters for the the system to decay into a stable stationary state.
However, we have seen in Sec.~\ref{sec4} that in the case of two degenerate modes there was a severe restriction that a stable steady state only exists for very weak optomechanical couplings, $G_{\theta}=\sqrt{2}G_{\Psi} < 2\sqrt{2}\gamma$.

After establishing the stability conditions, we now solve the set of the resulting equations of motion for the steady-state and find the following non-zero correlation functions
\begin{align}
\langle \delta b^{\dagger}\delta b\rangle &= \bar{n} +\frac{\left(\bar{n}+1\right)G_{\theta}^{2}}{2\left(U^{2} -G_{\theta}^{2}+4\gamma^{2}\right)} ,\nonumber\\
\langle \delta b \delta\Theta\rangle &= -\frac{i}{2}\frac{\left(\bar{n}+1\right) (2\gamma -iU)G_{\theta}}{\left(U^2-G_{\theta}^{2} +4\gamma^{2}\right)} ,\nonumber\\
\langle \delta \Theta^{\dagger}\delta\Theta\rangle &= \frac{(\bar{n}+1)G^{2}_{\theta}}{2\left(U^2 -G_{\theta}^{2}+4\gamma^{2}\right)} .\label{e60b}
\end{align}

If we now substitute Eq.~(\ref{e60b}) into Eq.~(\ref{eq39}), we arrive at the following expression for the anomalous cross correlation function
\begin{align}
\left|\eta_{(b,\theta)}\right|^{2} = \frac{\left(\bar{n}+1\right)\left(4\gamma^{2}+U^{2}\right)}{\left[2\bar{n}\left(4\gamma^{2}+U^{2}\right) -\left(\bar{n}-1\right)G_{\theta}^{2}\right]}  .\label{e60c}
\end{align}
The expression (\ref{e60c}) differs markedly from the one we encountered in Eq.~(\ref{eq70}), although they become identical in the limit of degenerate modes, when $U=0$. Similar to the case of degenerate modes, the condition for the violation of the Cauchy-Schwartz inequality, $|\eta_{(b,\theta)}|^{2}> 1$, is restricted to $\bar{n}<1$. However, in contrast to the case of degenerate modes, there is no restriction on $G_{\theta}$, as long as~$U>G_{\theta}$. For small $G_{\theta}$, the entanglement is almost insensitive to $U$. On the other hand, for large $G_{t}$, the entanglement depends crucially on the extend to which $G_{\theta}$ differs from $U$. When $G_{\theta}\approx U$, the correlation $|\eta_{(b,\theta)}|^{2}\approx 1$ indicating no entanglement between the modes, but $|\eta_{(b,\theta)}|^{2}\gg 1$ in the limit of $U\gg G_{\theta}$.
\begin{figure}[thb]
\begin{center}
\begin{tabular}{c}
\includegraphics[width=0.95\columnwidth]{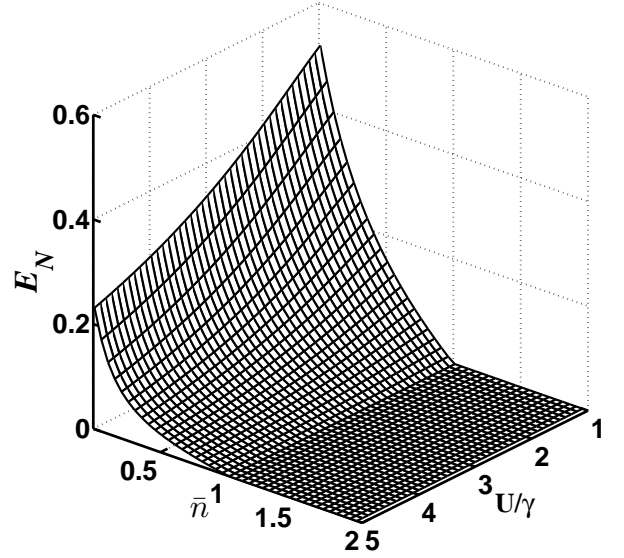}
\end{tabular}
\end{center}
\caption[nsa]{ \label{mirfig4} Variation of the logarithmic negativity $E_{N}$ with $\bar{n}$ and~$U/\gamma$ for $G_{\theta}=\gamma$.}
\end{figure}

These analysis are illustrated in Fig.~\ref{mirfig4}, which shows the logarithmic negativity for the case of the two colour entanglement calculated from the previously derived expressions, Eqs.~(\ref{e58e})-(\ref{e63b}), with the drift matrix ${\bf A}$ now given by
\begin{align}
\bf{A}& = \left(\begin{array}{cccc}
-\gamma&0&0&-\frac{1}{2}G_{\theta}\\
0&-\gamma&-\frac{1}{2}G_{\theta}&0\\
0&-\frac{1}{2}G_{\theta}&-\gamma&U\\
-\frac{1}{2}G_{\theta}&0&-U&-\gamma
\end{array}\right) .
\end{align}

It is seen that, as before for the degenerate case, an entanglement occurs for $\bar{n}<1$. However, in contrast to the degenerate case, whenever the Cauchy-Schwartz inequality is violated, the entanglement occurs over the entire range of $U$.

\subsection{Coupled three nondegenerate modes}

We now turn to the problem of determining conditions for correlations and entanglement between the three modes. We are particularly interested in the problem of entangling two independent modes that are simultaneously coupled to a third intermediate mode. This situation is encountered in Eq.~(\ref{e55b}) where two mutually independent and nondegenerate in frequency polaritons are simultaneously coupled to the mechanical mode. We note that both polaritons are coupled to the mechanical mode through a parametric interaction that can create an entanglement between the polaritons and the mechanical mode. An interesting question then arises whether this kind of the interaction could result in an entanglement between the polaritons. We solve Eq.~(\ref{e55b}) for the steady-state and see whether the anomalous cross correlation function $\langle\delta\Theta\delta\Pi\rangle$, necessary for entanglement between the polaritons, is different from zero.

To keep the mathematical complications to a minimum, we will take $\phi +\varphi =\pi/4$ so that the effective polaritons $\delta\Theta$ and $\delta\Pi$ are in the maximally entangled states. In this case, $G_{\theta}=-G_{\pi}\equiv G_{t}$. The situation of $\phi +\varphi =\pi/4$ holds for the case of the cavity field frequency on resonance with the exciton frequency $(\delta=0)$ and the coupling of the exciton mode to the cavity mode much stronger than the coupling of the mirror to the cavity mode, i.e. $\tilde{\Omega}\gg G_{q}$.

Let us first examine the stability conditions for the steady-state solutions of Eq.~(\ref{e55b}). It is not difficult to find that the eigenvalues of the $3\times3$ matrix, appearing in Eq.~(\ref{e55b}), are
\begin{align}
\lambda_{1} = \gamma ,\quad \lambda_{2,3} = \gamma \mp iB ,
\end{align}
where  $B= \sqrt{U^{2}-\frac{1}{2}G_{t}^{2}}$. Since $U\gg G_{t}/\sqrt{2}$, we have that the parameter $B$ is a real number for all values of the mechanical constant $G_{0}$. It is clear that the transformed operators are damped with the rate $\gamma$ independent of~$G_{0}$. In other words, there is no threshold for $G_{0}$, which means that the three-mode system will decay to a stabile steady-state independent of $G_{0}$.

If the mirror oscillates in a thermal field we then obtain from Eq.~(\ref{e55b}) the following nonzero steady-state correlation functions
\begin{align}
\langle \delta b^\dag \delta b\rangle &= \bar{n} + \frac{\left(\bar{n}+1\right)\left[3G_{t}^{2}+8(B^{2}+\gamma^{2})\right]G_{t}^{2}}{D} ,\nonumber\\
\langle\delta\Theta^\dag \delta\Theta\rangle &= \langle\delta\Pi^\dag \delta\Pi\rangle = \frac{(\bar{n}+1)\!\left[3G_{t}^{2}+8\left(B^2+\gamma^{2}\right)\right]\!G_{t}^{2}}{2D} ,\nonumber\\
\langle\delta\Theta \delta b\rangle &= \langle\delta\Pi \delta b\rangle^{\ast} \nonumber\\
=& \frac{i(\bar{n}\!+\!1)(\gamma\!+\!iU)[3G^2_{t}\!+\!4(B^{2}\!-\!2\gamma^{2})\!+\!12i\gamma U]G_{t}}{D} ,\nonumber\\
\langle\delta\Theta^\dag \delta\Pi\rangle &= \frac{(\bar{n}\!+\!1)\!\left[3G^2_{t} +4(B^{2}\!-\!2\gamma^{2})\!-\!12i\gamma U\right]\!G_{t}^{2}}{2D} ,\label{eq80}
\end{align}
with $D=8(B^2+\gamma^{2})(B^2+4\gamma^{2})$.

Unfortunately, the anomalous cross correlation function $\langle\delta\Theta\delta\Pi\rangle$ that is necessary for entanglement between the polaritons, is equal to~zero. Hence, the polaritons remain uncorrelated and therefore cannot be entangled. In that case, the indirect two-photon coupling between the polaritons which is provided by the oscillating mirror is effectively zero. The reason for it is that the oscillating mirror couples to the polaritons with the opposite phases, as it is evident from Eq.~(\ref{e55b}). In physical terms, the oscillating mirror establishes the phase difference $\phi_{\theta}-\phi_{\pi}$ rather than the phase sum $\phi_{\theta}+\phi_{\pi}$ between the two polaritons. In this case, that mutual behaviour of the polaritons and the mirror does not create two-photon correlations between the polaritons.

In order to gain some appreciation of the magnitude of the first-order coherence, we evaluate the degree of the first-order coherence between the polaritons, we find
\begin{align}
\left|\gamma_{(\theta,\pi)}\right| = \frac{\left|3G^2_{t} +4(B^{2}-2\gamma^{2})+12i\gamma U\right|}{3G_{t}^{2}+8(B^2+\gamma^{2})} .\label{eq82}
\end{align}
Thus we see that the oscillating mirror, although not being able to entangle the polaritons, it in turn makes the polaritons partly coherent. Recall that the polariton modes are in thermal states, $g^{(2)}_{\theta} =g^{(2)}_{\pi}=2$. Thus, two independent thermal modes can be made by the oscillating mirror mutually coherent and the degree of coherence can, in principle, be as large as unity.
Note that $|\gamma_{(\theta,\pi)}|$ is independent of $\bar{n}$. The reason for this is that the coherence depends on the phase relation between the modes and the information about the phase is not carried out by the thermal fluctuations. Moreover, the expression~(\ref{eq82}) shows that the polaritons are not perfectly coherent, the coherence raises with the coupling strength $G_{t}$ and becomes unity only in the limit of $G_{t}\gg \tilde{\Omega},\gamma$.

Since the anomalous cross correlations were not generated when the mirror was oscillating in the thermal field, we now turn to evaluate the cross correlation function $\langle\delta\Theta\delta\Pi\rangle$ assuming that the mirror oscillates in a squeezed vacuum. We have already seen that in the two-mode case, the coupling the mirror to a squeezed field resulted in the generation of the correlation functions that were zero in the thermal field. Following this observation, we recalculate $\langle\delta\Theta\delta\Pi\rangle$ using Eq.~(\ref{eq72a}), and find that in the squeezed field the anomalous cross correlation function is now different from zero and is of the form
\begin{eqnarray}
\langle\delta\Theta\delta\Pi\rangle = \frac{mG^2_{t}}{2D}\left[3G_{t}^{2}+8\left(B^2+\gamma^{2}\right)\right] .\label{eq84}
\end{eqnarray}
The correlation functions determining the occupation of the polariton modes $\langle\delta\Theta^{\dagger}\delta\Theta\rangle$ and $\langle\delta\Pi^{\dagger}\delta\Pi\rangle$ are not changed by the squeezed vacuum and are identical with the results found for the thermal field, Eqs.~(\ref{eq80}). This leads to a quite simple expression for the degree of the anomalous cross correlations
\begin{align}
\left|\eta_{(\theta,\pi)}\right| = \frac{|m|}{\bar{n}+1} .\label{eq85}
\end{align}
This expression indicates that the polaritons could be  entangled as long as $|m|\neq 0$.
We note that $|\eta_{(\theta,\pi)}|$, which is necessary for the generation of entanglement between the polaritons, is independent of the cavity and the mirror parameters. It is solely determined by the parameters of the squeezed vacuum field in which the mirror oscillates. The reason for it is in the broadband nature of the squeezed vacuum, we have assumed here, that all modes of the vacuum are equally squeezed independent of frequency.

It should be stressed that the squeezed vacuum creates not only the anomalous cross correlation between the modes but also creates the anomalous autocorrelations in the modes. A straightforward calculation shows that
\begin{align}
\langle \delta\Theta^2\rangle = \langle \delta\Pi^2\rangle = \frac{mG^2_{t}}{2D}\left[3G^2_{t}\!+\!4(B^{2}\!-\!2\gamma^{2})\!+\!12i\gamma U\right] ,\label{eq86}
\end{align}
and then the degree of anomalous autocorrelations are
\begin{align}
\left|\eta_{(\theta,\theta)}\right| = \left|\eta_{(\pi,\pi)}\right| = \frac{|m|}{\bar{n}+1}\frac{\left|3G^2_{t} +4(B^{2}-2\gamma^{2})\!+\!12i\gamma U\right|}{3G_{t}^{2}+8(B^2+\gamma^{2})} .\label{eq87}
\end{align}
Hence, the creation of the anomalous cross correlations between the polaritons is accompanied by the creation of the anomalous autocorrelations in the modes. It is easily verified that the inequality $|\eta_{(\theta,\pi)}|^{2} >|\eta_{(\theta,\theta)}||\eta_{(\pi,\pi)}|=|\eta_{(\theta,\theta)}|^{2}$ is always satisfied regardless of the value of $G_{t}$.

Normally, we would expect that the requirement for the anomalous cross correlations to satisfy the inequality $|\eta_{(\theta,\pi)}|^{2} >|\eta_{(\theta,\theta)}||\eta_{(\pi,\pi)}|$ is sufficient for entanglement between the modes. This was suggested by the inequalities~(\ref{eq70}) and~(\ref{e60c}) which were sufficient conditions for entanglement between one of the polaritons and the mechanical mode. However, substituting Eqs.~(\ref{eq82}), (\ref{eq85}) and (\ref{eq87}) into Eq.~(\ref{eq62}), we find that the Cauchy-Schwartz parameter is of the form
\begin{align}
\chi_{(\theta,\pi)} = \left[1 +\frac{\left(1-|\Upsilon|^{2}\right)\left(1-\frac{|m|^{2}}{\left(\bar{n}+1\right)^{2}}\right)}{1+|\Upsilon|^{2}+\frac{|m|^{2}}{(\bar{n}+1)^{2}}}\right]^{2} .
\end{align}
where
\begin{align}
|\Upsilon|^{2} = 1 - \frac{24(G_{t}^{2}+2B^{2})(B^2+\gamma^{2})}{\left[3G_{t}^{2}+8(B^2+\gamma^{2})\right]^{2}} .\label{eq89}
\end{align}
Since $|m|^{2}=\bar{n}(\bar{n}+1)<(\bar{n}+1)^{2}$ and $|\Upsilon|^{2}\leq 1$, the Cauchy-Schwartz inequality cannot be violated. We therefore conclude that the presence of a strong first-order coherence between the polariton modes prevents the modes from being entangled.

We may summarize that the parametric interaction between the polaritons and the mechanical mode rules out the creation of entanglement between the polaritons. The effect of these simultaneous parametric interactions is to create the first-order coherence between the polaritons.

Consider now a different scenario. Suppose that one of the polaritons is coupled to the mirror by a parametric interaction, but the other one is coupled by a linear-mixing interaction. We shall demonstrate that this kind of coupling can occur in our system and it turns out is sufficient to create entanglement between two modes that are not directly coupled to each other. For example, if instead of the two polaritons, we consider their linear superpositions
\begin{align}
\delta A_{1} = \frac{1}{\sqrt{2}}\left(\delta\Theta -\delta\Pi\right) ,\quad \delta A_{2} = \frac{1}{\sqrt{2}}\left(\delta\Theta +\delta\Pi\right) ,
\end{align}
we readily find that Eqs.~(\ref{e55b}) can be transformed into equations
\begin{align}
\delta\dot{\tilde b}^{\dag} &= -\gamma \delta{\tilde b}^{\dag} +\frac{i}{\sqrt{2}}G_{t}\delta A_{1} + \sqrt{2\gamma}\,\tilde{\xi}^{\dag}(t)  ,\nonumber\\
\delta\dot{A}_{1} &= -\gamma\delta A_{1} -iU\delta A_{2} -\frac{i}{\sqrt{2}}G_{t}\delta{\tilde b}^{\dag} + \sqrt{2\gamma}\, A_{1}^{in}(t)  ,\nonumber\\
\delta\dot{A}_{2} &= -\gamma\delta A_{2} - iU\delta A_{1} + \sqrt{2\gamma}\, A_{2}^{in}(t) .\label{eq93}
\end{align}
It follows that in this case not $\delta b^{\dagger}$ but $\delta A_{1}$ plays the role of the intermediate mode, the mechanical mode $\delta b^{\dagger}$ is coupled to $\delta A_{1}$ by the parametric amplification process, and the mode~$\delta A_{2}$ is coupled to $\delta A_{1}$ by the linear mixing process. There is no direct coupling between the modes $\delta b^{\dagger}$ and~$\delta A_{2}$. Nevertheless, we shall demonstrate that the modes $\delta b^{\dagger}$ and $\delta A_{2}$ can be entangled. In order to illustrate the idea, we proceed to determine the anomalous cross correlation function $\langle \delta A_2\delta b\rangle$ from the equations of motion (\ref{eq93}). With the help of Eqs.~(\ref{e22}) and~(\ref{e23}), we solve Eqs.~(\ref{eq93}) for the steady-state and find the following nonzero correlation functions
\begin{align}
\langle \delta b^{\dagger} \delta b\rangle &= \bar{n} +\frac{\left(\bar{n}+1\right)\left[3G_{t}^{2}+8(B^{2}+\gamma^{2})\right]G^{2}_t}{D} ,\nonumber\\
\langle \delta A_{1}\delta b\rangle &= \frac{-2 i\gamma\left(\bar{n}+1\right)\left[3G_{t}^{2}+8(B^{2}+\gamma^{2})\right]G_{t}}{D} ,\nonumber\\
\langle \delta A_2\delta b\rangle &= -\frac{\sqrt{2}(\bar{n}+1)\left[3G_{t}^{2}+4(B^{2}+\gamma^{2})\right]U G_t}
{ D} ,\nonumber\\
\langle \delta A^\dagger_1\delta A_{1}\rangle &=\frac{\left(\bar{n}+1\right)\left(B^2+4\gamma^{2}-\frac{1}{2}G^{2}_t\right)G^{2}_t}{D} ,\nonumber\\
\langle \delta A^\dagger_1\delta A_{2}\rangle &= -\frac{6i\left(\bar{n}+1\right)\gamma UG^{2}_t}{D},\nonumber\\
\langle \delta A_{2}^{\dagger}\delta A_2\rangle &= \frac{6\left(\bar{n}+1\right)U^{2}G^{2}_t}{D} .\label{e98}
\end{align}
It is easy to see that the anomalous cross correlation function $\langle \delta A_2\delta b\rangle$ is nonzero indicating that the combination of the parametric and linear-mixing interactions  between three modes may result in entanglement between two modes that are not directly coupled to each other. Because the anomalous autocorrelations and the first-order coherence are zero, $|\eta_{(A_{2},A_{2})}|^{2}=|\eta_{(b,b)}|^{2}=|\gamma_{(A_{2},b)}|^{2}=0$, the sufficient condition for entanglement between the modes~$\delta b$ and $\delta A_{2}$ is that $|\eta_{(A_{2},b)}|^{2}$ satisfies the inequality $|\eta_{(A_{2},b)}|^{2}>1$.
\begin{figure}[thb]
\begin{center}
\begin{tabular}{c}
\includegraphics[width=0.95\columnwidth]{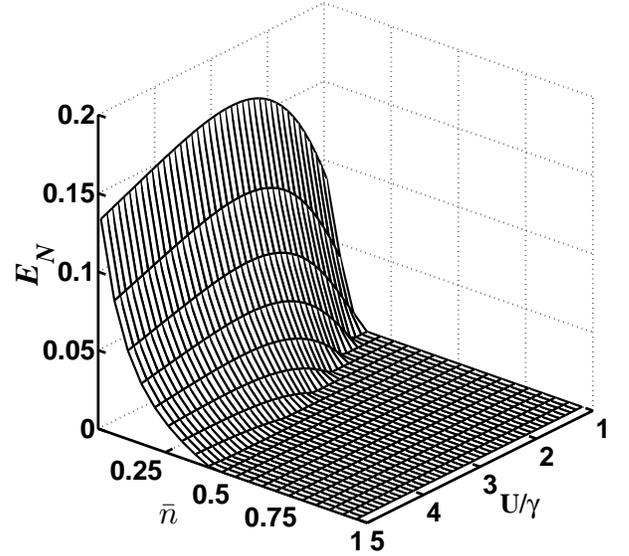}
\end{tabular}
\end{center}
\caption[nsa]{ \label{mirfig5} Variation of the logarithmic negativity $E_{N}$ with $\bar{n}$ and~$U$ illustrating  the entanglement creation between the indirectly coupled modes $\delta A_{2}$ and $\delta b$ for $G_{t}=\gamma$.}
\end{figure}

Figure~\ref{mirfig5} shows the logarithmic negativity $E_{N}$ for fixed $G_{t}=\gamma$ and gradually increasing $\bar{n}$ and $U$. As in the case of two coupled modes, the logarithmic negativity is determined from the steady-state correlation matrix, which in the present case is a $6\times 6$ matrix derived by solving an equation that is formally identical with Eq.~(\ref{e61e}), but with the drift matrix ${\bf A}$ given by
\begin{align}
\bf{A}& = \left(\begin{array}{cccccc}
-\gamma&0&0&-\frac{G_{t}}{\sqrt{2}}&0&0\\
0&-\gamma&-\frac{G_{t}}{\sqrt{2}}&0&0&0\\
0&-\frac{G_{t}}{\sqrt{2}}&-\gamma&0&0&U\\
-\frac{G_{t}}{\sqrt{2}}&0&0&-\gamma&-U&0\\
0&0&0&U&-\gamma&0\\
0&0&-U&0&0&-\gamma
\end{array}\right) .
\end{align}
and with the diffusion matrix ${\bf D} =\textrm{diag}[(2\bar{n}+1)\gamma,(2\bar{n}+1)\gamma,\gamma,\gamma,\gamma,\gamma]$.

It can be seen from Fig.~\ref{mirfig5} that entanglement occurs for $\bar{n}<0.5$ where it is present for all $U$. The degree of the entanglement increases with~$U$, the strength of the linear mixed process. Note that the entanglement is reduced when $U$ is comparable to $G_{t}$, and increases as $U$ departures from $G_{t}$.

We conclude this section that two independent modes may become entangled by a suitable coupling to an intermediate mode. The entanglement occurs when one of the modes couples to the intermediate mode by the parametric interaction process whereas the other couples by the linear-mixing interaction.

\section{Experimental considerations}~\label{sec6}

Finally, we examine parameter ranges in which the predicted coherence and entanglement effects could be observed with the current experiments.
As a check on the validity of the limitation to only the $k=1$ exciton, we evaluate the formula~(\ref{e20}) for $f_{k}$ using experimentally realistic parameters of an optical lattice composed of $^{85}$Rb atoms~\cite{zr07,zr09}. By taking $N=10^3$ sites, the cavity mode volume $V =10^{-10}$ [m$^{3}$], the atomic transition dipole moment $\mu= 5 \times10^{-29}$ [Cm] of the hyperfine transition 5 $^{2}S_{1/2} -$ 5 $^{2}P_{3/2}$ in an $^{85}$Rb atom, and the cavity frequency $\omega_{c}$ on resonance with the first $k=1$ exciton mode that is comparable to the atomic transition frequency of $\omega_a =2.5\times10^{15}$ [Hz], we obtain for the coupling strength of the first $k=1$ exciton, $f_{1}/\hbar = 1.6\times10^8$ [Hz], and for the third $k=3$ exciton, $f_{3}/\hbar = 5.3\times10^7$ [Hz] that is one order smaller than the $k=1$ coupling strength. This simple estimation of the coupling strength shows that $f_{k}$ decays very fast with $k$. Thus, our theory can describe the single exciton system quite accurately.

The stability condition for the three mode coupling case requires $U > G_{t}/\sqrt{2}$, where $U=\sqrt{\tilde{\Omega}^{2} + G_{q}^{2}}$ and $G_{t}\approx G_{0}$. Since $\tilde{\Omega}\sim |f_{1}|/\hbar = 1.6\times10^8$ [Hz] and the typical coupling strengths $G_{0}/2\pi\approx 10^{6}$ [Hz], the inequality $U > G_{t}/\sqrt{2}$ can be satisfied with the realistic experimental parameters.

\section{Summary}~\label{sec7}

In summary, we have presented an analytical study of coherence and correlation effects produced in a single-mode nano-mechanical cavity containing an optical lattice of regularly trapped atoms. The system considered is equivalent to a three-mode system composed of two polariton modes and one mechanical mode. We have shown that the system is capable of generating a wide class of coherence and correlation effects, ranging from the first-order coherence, the anomalous autocorrelations and anomalous cross correlations between the modes.
We have been particularly interested in the relationship between the generation of entanglement and the first-order coherence in the system. The results show that the generation of the first-order coherence between two modes of the system is equally effective in destroying entanglement between these modes. There is no entanglement between the independent polariton modes when both modes are simultaneously coupled to the mechanical mode by the parametric (squeezing-type) interaction. There is no entanglement between the polaritons even if the oscillating mirror is damped by a squeezed vacuum field. The intermediate mechanical mode effectively creates the first-order coherence between the modes. Finally, we have shown that in order to effectively entangle independent modes, in this system, one of the modes should be coupled to the intermediate mode by a parametric interaction but the other mode should be coupled by the linear-mixing (beamsplitter-type) interaction.

\section*{Acknowledgment}

This work was supported by the National Natural Science Foundation of China (Grant No. 11074087), , and the Natural Science Foundation of Hubei Province (Grant No. 2010CDA075), the Nature Science Foundation of Wuhan City (Grant No. 201150530149), and the National Basic Research Program of China(Grant No. 2012CB921602).


\begin{thebibliography}{99}

\frenchspacing

\bibitem{gm09} C. Genes, A. Mari, D. Vitali, and P. Tombesi, Adv. At. Mol. Opt. Phys. {\bf 57}, 33 (2009).

\bibitem{agh10} M. Aspelmeyer, S. Gr\"{o}blacher, K. Hammerer, and N. Kiesel, J. Opt. Soc. Am. B {\bf 27}, A189 (2010).

\bibitem{td09} J. D. Teufel, T. Donner, M. A. Castellanos-Beltran, J. W. Harlow, and K. W. Lehnert, Nature Nanotechnology {\bf 4}, 820 (2009).

\bibitem{gh09} S. Gr\"{o}blacher, K. Hammerer, M. R. Vanner, and M. Aspelmeyer, Nature {\bf 460}, 724 (2009).

\bibitem{fg06} A. Ferreira, A. Guerreiro, and V. Vedral, Phys. Rev. Lett. {\bf 96}, 060407 (2006).

\bibitem{vg07} D. Vitali, S. Gigan, A. Ferreira, H. R. B\"{o}hm, P. Tombesi, A. Guerreiro, V. Vedral, A. Zeilinger, and M. Aspelmeyer, Phys. Rev. Lett. {\bf 98}, 030405 (2007).

\bibitem{pv07} M. Paternostro, D. Vitali, S. Gigan, M. S. Kim, C. Brukner, J. Eisert, and M. Aspelmeyer, Phys. Rev. Lett. {\bf 99}, 250401 (2007).

\bibitem{vt07} D. Vitali, P. Tombesi, M. J. Woolley, A. C. Doherty, and G. J. Milburn, Phys. Rev. A {\bf 76}, 042336 (2007).

\bibitem{bg08} M. Bhattacharya, P. L. Giscard, and P. Meystre, Phys. Rev. A {\bf 77}, 013827 (2008).

\bibitem{gm08} C. Genes, A. Mari, P. Tombesi, and D. Vitali, Phys. Rev. A {\bf 78}, 032316 (2008).

\bibitem{ig08} H. Ian, Z. R. Gong, Y.-X. Liu, C. P. Sun, and F. Nori, Phys. Rev. A {\bf 78}, 013824 (2008).

\bibitem{pc10} M. Paternostro, G. De Chiara, and G. M. Palma, Phys. Rev. Lett. {\bf 104}, 243602 (2010).

\bibitem{cp11} G. De Chiara, M. Paternostro, and G. M. Palma, Phys. Rev. {\bf 83}, 052324 (2011).

\bibitem{mg02} S. Mancini, V. Giovannetti, D. Vitali, and P. Tombesi, Phys. Rev. Lett. {\bf 88}, 120401 (2002).

\bibitem{hp08} M. J. Hartmann and M. B. Plenio, Phys. Rev. Lett. {\bf 101}, 200503 (2008).

\bibitem{hw09} K. Hammerer, M. Wallquist, C. Genes, M. Ludwig, F. Marquardt, P. Treutlein, P. Zoller, J. Ye, and H. J. Kimble, Phys. Rev. Lett. {\bf 103}, 063005 (2009).

\bibitem{wh10} M. Wallquist, K. Hammerer, P. Zoller, C. Genes, M. Ludwig, F. Marquardt, P. Treutlein, J. Ye, and H. J. Kimble, Phys. Rev. A {\bf 81}, 023816 (2010).

\bibitem{ck11} S. Camerer, M. Korppi, A. J\"{o}ckel, D. Hunger, T. W. H\"{a}nsch, and P. Treutlein, Phys. Rev. Lett. {\bf 107}, 223001 (2011).

\bibitem{jh09} J. D. Jost, J. P. Home, J. M. Amini, D. Hanneke, R. Ozeri, C. Langer, J. J. Bollinger, D. Leibfried, and D. J. Wineland, Nature {\bf 459}, 683 (2009).

\bibitem{bv11} Sh. Barzanjeh, D. Vitali, P. Tombesi, and G. J. Milburn, Phys. Rev. A {\bf 84}, 042342 (2011).

\bibitem{ak10} U. Akram, N. Kiesel, M. Aspelmeyer, and G. J. Milburn, New J. Phys. {\bf 12}, 083030 (2010).

\bibitem{hw11} S. G. Hofer, W. Wieczorek, M. Aspelmeyer, and K. Hammerer, Phys. Rev. A {\bf 84}, 052327 (2011).

\bibitem{zh11} L. Zhou, Y. Han, J. Jing, and W. Zhang, Phys. Rev. A {\bf 83}, 052117 (2011).

\bibitem{gh86} R. Ghosh, C. K. Hong, Z. Y. Ou, and L. Mandel, Phys. Rev. A {\bf 34}, 3962 (1986).

\bibitem{ow90} Z. Y. Ou, L. J. Wang, X. Y. Zou, and L. Mandel, Phys. Rev. A {\bf 41}, 1597 (1990).

\bibitem{mg96} C. H. Monken, A. Garuccio, D. Branning, J. R. Torgerson, F. Narducci, and L. Mandel, Phys. Rev. A {\bf 53}, 1782 (1996).

\bibitem{zr09} H. Zoubi and H. Ritsch, EPL {\bf 87}, 23001 (2009).

\bibitem{zr07} H. Zoubi and H. Ritsch, Phys. Rev. A {\bf 76}, 013817 (2007).

\bibitem{jb98} D. Jaksch, C. Bruder, J. I. Cirac, C. W. Gardiner, and P. Zoller, Phys. Rev. Lett. {\bf 81}, 3108 (1998).

\bibitem{mm06} D. Meiser and P. Meystre, Phys. Rev. A {\bf 73}, 033417 (2006).

\bibitem{hs10} K. Hammerer, K. Stannigel, C. Genes, P. Zoller, P. Treutlein, S. Camerer, D. Hunger, and T. W. H\"{a}nsch, Phys. Rev. A {\bf 82}, 021803(R) (2010).

\bibitem{bp03} S. L. Braunstein and A. K. Pati, {\it Quantum Information Theory with Continuous Variables}, (Kluwer, Dordrecht, 2003).

\bibitem{wn92} C. Weisbuch, M. Nishioka, A. Ishikawa, and Y. Arakawa, Phys. Rev. Lett. {\bf 69}, 3314 (1992).

\bibitem{li05} G. X. Li, S. P. Wu, and G. M. Huang, Phys. Rev. A {\bf 71}, 063817 (2005).

\bibitem{li07} G. X. Li, H. T. Tan, and M. Macovei, Phys. Rev. A {\bf 76}, 053827 (2007).

\bibitem{mw95} L. Mandel and E. Wolf,  {\it Optical Coherence and Quantum Optics} (Cambridge: Cambridge University Press, 1995).

\bibitem{ks80} A. P. Kazantsev, V. S. Smirnov, and V. P. Sokolov, Optics Commun.  {\bf 35}, 209 (1980).

\bibitem{lo84} M. J. Collett, D. F. Walls, and P. Zoller, Optics Commun. {\bf 52}, 145 (1984).

\bibitem{ag86} G. S. Agarwal, Phys. Rev. A {\bf 33}, 2472 (1986).

\bibitem{hr87} A. Heidmann and S. Reynaud, J. Mod. Opt. {\bf 34}, 923 (1987).

\bibitem{ft88} Z. Ficek and R. Tana\'s, Z. Phys. D {\bf 9}, 27 (1988).

\bibitem{sl11} L. H. Sun, G. X. Li, W. J. Gu, and Z. Ficek, New J. Phys. {\bf 13}, 093019 (2011).

\bibitem{bk30} P. Meystre and M. Sargent, {\it Elements of Quantum Optics} (Springer, New York, 1999), p. 330.

\bibitem{si00} R. Simon, Phys. Rev. Lett. {\bf 84}, 2726 (2000).

\bibitem{as04} G. Adesso, A. Serafini, and F. Illuminati, Phys. Rev. A {\bf 70}, 022318 (2004).

\bibitem{df04} {\it Quantum Squeezing}, eds. P. D. Drummond and Z. Ficek, (Springer, New York, 2004).

\bibitem{vc05} A. S. Villar, L. S. Cruz, K. N. Cassemiro, M. Martinelli, and P. Nussenzveig, Phys. Rev. Lett. {\bf 95}, 243603 (2005).

\bibitem{cb10} A. S. Coelho, F. A. S. Barbosa, K. N. Cassemiro, A. S. Villar, M. Martinelli, and P. Nussenzveig, Science {\bf 326}, 823 (2010).

\bibitem{wg10} M. F. Wang, W. J. Gu, Q. L. Jin, and Y. Z. Zheng, Phys. Rev. A {\bf 82}, 042323 (2010).

\bibitem{lg10} Y. Li, X. Guo, Z. Bai, and C. Liu, Appl. Phys. Lett. {\bf 97}, 031107 (2010).

\bibitem{sl10} A. Samblowski, C. E. Lauk\"{o}tter, N. Grosse, P. K. Lam, and R. Schnabel, AIP Conf. Proc. {\bf 1363}, 219 (2011).

\bibitem{co11} D. Cuozzo and G. L. Oppo, Phys. Rev. A {\bf 84}, 043810 (2011).

\bibitem{tan11} H. T. Tan and G. X. Li, Phys. Rev. A {\bf 84}, 024301 (2011).

\end{thebibliography}
\end{document}